**A Novel Multi-Target Approach for simultaneous enhancement of brightness, intensity, and bi-spectral performance of cold neutron sources**

**A. Ioffe[1], P. Konik[2,3,*], D. Shapiro[4]**

**1** Department of Physics and Technology, Al-Farabi Kazakh National University, Almaty, Kazakhstan
**2** Spallation Neutron Source Science Center, Dongguan, China
**3** Institute of High Energy Physics, Chinese Academy of Sciences, Beijing, China
**4** Budapest Neutron Centre, HUN-REN Centre for Energy Research, Budapest, Hungary

***Corresponding author: konikpetr@ihep.ac.cn**

## Abstract

Increasing the brightness of cold neutron beams is a central objective in modern cold neutron source development, since it largely governs the sample intensity for high-resolution neutron instruments. Raising accelerator beam power is constrained by the target heat-removal limit of 0.5–1 kW/cm², and increasing the moderator length beyond about 20–30 cm becomes inefficient owing to the finite mean free path of cold neutrons. We propose reversing the conventional design logic: rather than tailoring the cryogenic moderator to a fixed target–premoderator source, we tailor the thermal-neutron source geometry to a high-performance moderator configuration.

The concept is realized as a staircase moderator assembly of several shorter compressed moderators, each with its own target–premoderator unit, which distributes neutron production among multiple units and thus enables higher accelerator powers while preserving the brightness advantages of compressed moderators. This approach can provide additional brightness and intensity gains relative to a single low-dimensional or conventional voluminous moderator at the desired transverse beam size and may naturally generate bi-spectral neutron beams, thereby extending the dynamic range of neutron instruments.

## 1. Introduction

One of the main directions in the development of modern cold neutron sources is to increase the brightness of the emerging neutron beams, since brightness largely determines the intensity at the sample for most high-resolution neutron instruments, which require strong collimation of the incident beam [1].

Currently, low-dimensional (quasi-2D or quasi-1D) parahydrogen (p-$H_2$) moderators are considered among the most effective means of increasing neutron brightness [2]. Such moderators are planned for use at new neutron sources under construction or in the design phase, including ESS [3], STS [4], and HBS [5]. In particular, compressing a p-$H_2$ moderator from a 3D geometry with dimensions of $(10 \times 10 \times 10)$ cm$^3$ $(W \times H \times L)$ to a quasi-2D geometry of $(10 \times 3 \times 10)$ cm$^3$ by reducing its height $H$ from 10 cm to 3 cm, yields an approximately 2.5-fold increase in brightness. This approach is implemented, for example, in the ESS moderator design [3].

However, compressing a p-$H_2$ moderator leads to a decrease in the intensity of the emerging neutron beam. For a moderator height of approximately 3 cm, the beam intensity is already about 20% lower than that obtained with a 10-cm-high moderator [3]. Although further reducing the moderator height, for example to 1 cm, increases the brightness gain due to geometric compression to approximately a factor of five, this comes at the cost of an intensity loss of about 50%. At the same time, the relatively large transverse beam dimension can be exploited to implement additional neutron focusing in the direction perpendicular to the scattering plane at the sample, providing an additional means of enhancing the neutron flux at the sample.

Furthermore, in our earlier work [6], we showed that, for any given instrument and any specified distance between the moderator and the neutron-guide entrance, determined by the proton-beam power incident on the target, there is an optimum moderator size that provides a balance between moderator brightness and its intensity. Beyond this optimum, additional compression reduces the useful neutron flux at the sample

Thus, reducing the moderator size below this optimum is not a rational approach to increasing the neutron intensity delivered to the instrument.

A straightforward way to increase this flux is to raise the accelerator beam power and thus the neutron flux incident on the cryogenic moderator. The maximum usable beam power, however, is limited by the target's heat-removal capacity and its ability to withstand the associated thermal and mechanical stresses (e.g., heat loads of 0.5–1 $kW/cm^2$ for such High-Current Accelerator-Driven Neutron Sources (HiCANS) as HBS [7]). Therefore, within this approach, increasing the moderator length with a given moderator face is the only way to accommodate higher accelerator beam power. However, as will be shown in this work, owing to the finite mean free path of cold neutrons, further increasing the moderator length beyond approximately 20–30 cm, while proportionally increasing the accelerator beam power, becomes inefficient.

In the present work, we propose a new approach: rather than optimizing the cryogenic moderator geometry for a fixed thermal-neutron source geometry comprising the target and premoderator, we adapt the geometry of the thermal-neutron source itself to the selected high-performance cryogenic moderator configuration. The configuration considered here is a staircase moderator assembly consisting of several shorter compressed moderators [8], each paired with its own target–premoderator unit. Such a configuration offers a practical route to exploiting higher accelerator powers by distributing neutron production among multiple moderator units while retaining the brightness advantages of compressed moderators.

It will be shown that the use of an appropriately designed multi-target configuration makes it possible to achieve additional gains in both brightness and intensity compared with a single low-dimensional or conventional voluminous moderator, while maintaining the desired transverse beam size [6]. Furthermore, such a configuration may naturally produce bi-spectral neutron beams, allowing to extend dynamic range of the neutron instruments.

## 2. Low-dimensional para-hydrogen moderators

The operating principle of low-dimensional p-$H_2$ moderators [9] relies on the features of neutron interaction with p-$H_2$. On one hand, thermal neutrons with a typical energy of $E \approx 25$ meV are scattered by p-$H_2$ molecules, losing an energy of $\Delta E \approx 14.7$ meV by inducing the transition of p-$H_2$ to the ortho state (para-to-ortho conversion) and effectively becoming cold in a single collision within approximately 2 cm of the moderator's surface (see Fig. 1b). This means that the conventional moderator's volume required for a neutron to scatter several times before slowing down is not necessary, and the moderator can be compressed (so-called low-dimensional or flat moderator). On the other hand, cold neutron's cross-section for p-$H_2$ is much smaller than that of thermal neutrons (see Fig. 1a), so cold neutrons can be extracted from a large depth, and the moderator's length can become significant. Both effects result in an enhanced brightness in comparison with a voluminous moderator.

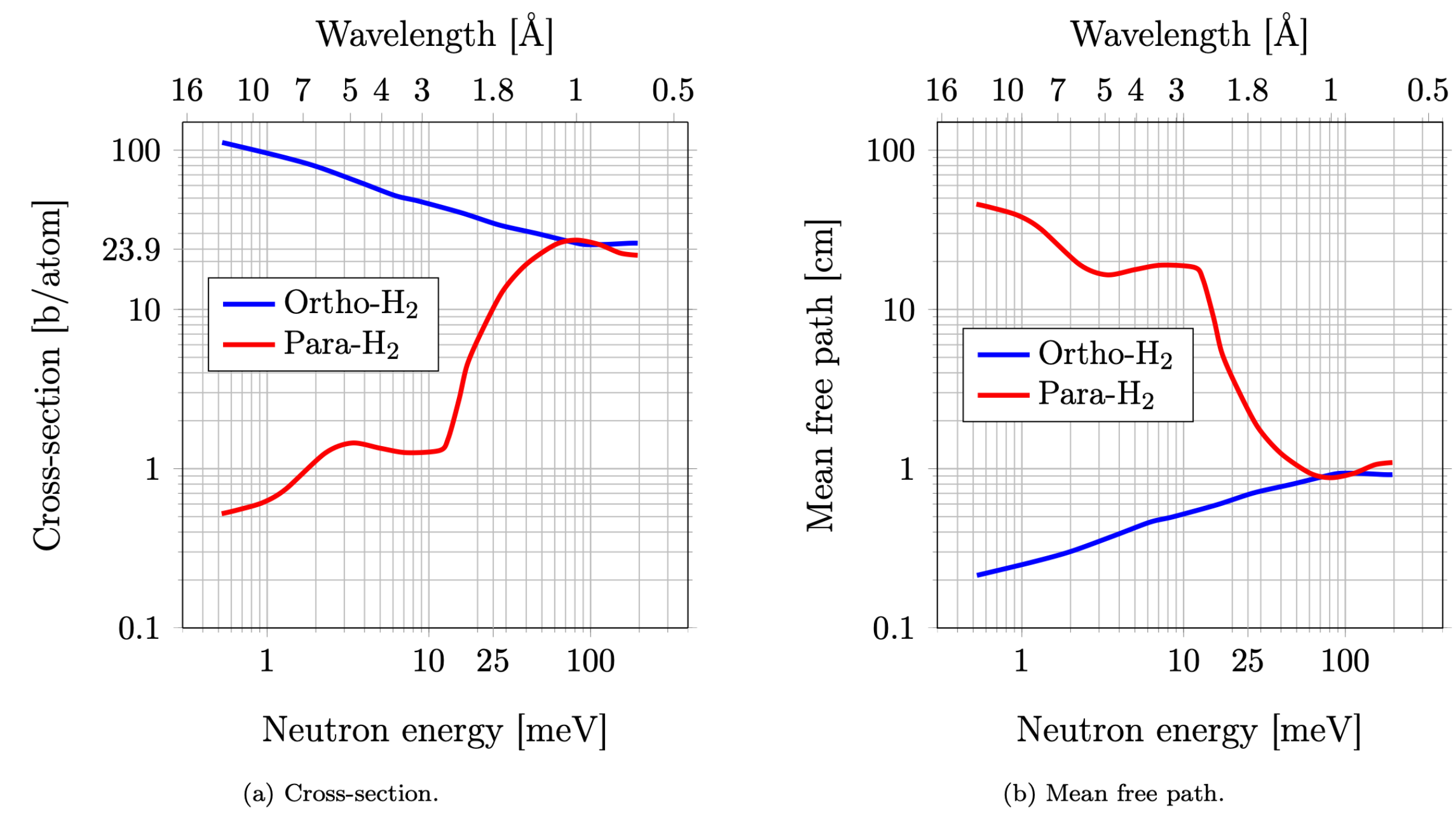

**Fig. 1.** Energy dependence of the neutron interaction cross section in ortho- and para-hydrogen at 20 K. Adapted from [8] based on the data from [10, Fig. 30, p. 367].

### 2.1. Selection of the para-hydrogen moderator length

Consider a p-$H_2$ moderator (1) of length $L$ and cross-sectional area ($W \times H$) (Fig. 2), one of whose horizontal surfaces is uniformly illuminated by thermal neutrons over an area ($W \times L$). These neutrons originate from a thermal moderator (premoderator) (2), in which fast neutrons with energies of a few MeV, produced in the target irradiated by a proton beam, are slowed down to thermal energies (25–100 meV). The mean free path of thermal neutrons $\Lambda_{th}(E)$ with energies $E$ > 50 meV in p-$H_2$ is less than 1 cm (and less than 2.5 cm for neutrons with $E$ > 25 meV; see Fig. 1). Consequently, their density within the moderator decreases rapidly with distance from the illuminated surface.

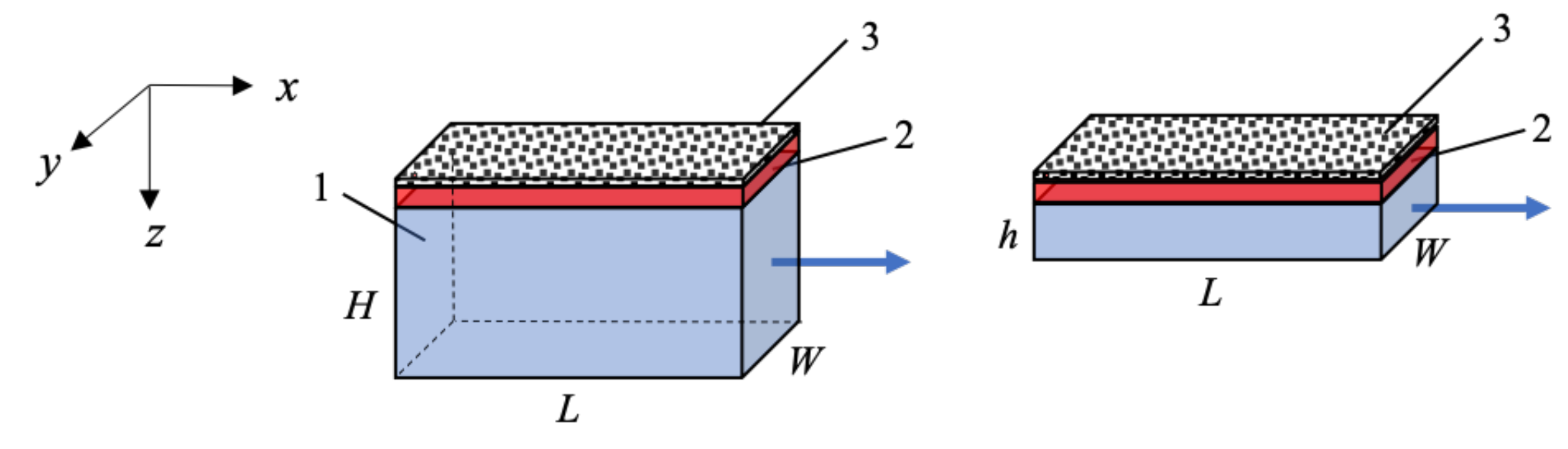

Fig. 2. Para-hydrogen moderator illuminated by thermal neutrons from a target–premoderator unit. Left: 3D moderator; right: compressed quasi-2D moderator ($h$ <$H$). 1– p-$H_2$ moderator; 2– thermal premoderator; 3–proton target. The blue arrows indicate the direction of cold-neutron emission.

Reducing the height $H$ of the 3D moderator to a smaller value $h$ (hereafter referred to as *compression*) decreases the intensity of the emitted cold neutron beam. However, the beam brightness, defined as the ratio of intensity to the area of the exit cross section (in the small-solid-angle approximation and at a fixed detector distance), increases substantially [9]. The brightness of a compressed moderator is typically compared with that of a voluminous moderator with a height of 10 cm, thereby defining the compression gain. As shown below in Fig. 6, this gain increases almost exponentially as the moderator height decreases.

In contrast to thermal neutrons, cold neutrons have a much longer mean free path in para-hydrogen, typically 10–20 cm (approximately 20 cm for neutrons with $E$ < 12 meV; see Fig. 1). Consequently, cold neutrons generated deep inside the moderator can escape efficiently through its end surface and contribute to the emerging beam. Thus, increasing the moderator length $L$ , together with the length of the thermal-neutron source, increases both the intensity and the brightness of the emerging beam.

For simplicity, in the following analysis, we assume that the moderator surface is illuminated by a uniform thermal-neutron flux Φ ($n/cm^2/s$), so that the total neutron rate (intensity) is given by the product of Φ and the illuminated area: $I_{total} = \Phi \cdot (W \times L)$. Consequently, increasing the moderator length $L$ requires a proportional increase in the proton beam intensity and, hence, in the accelerator power $P$.

In the single-collision approximation, when we assume that cold neutrons are produced via just one collision in p-$H_2$, the spectral intensity of the cold-neutron beam emitted along the longitudinal direction (Fig. 2) is expressed as a function of neutron energy $E_{out} = E - \Delta E$ as follows:

$$I_{out}(E_{out}, L, H, T) \sim M(E,T) \cdot \int_{-\pi/2}^{\pi/2} cos\,(\theta)\left\{1 - exp\left[-\frac{S(\theta,H,L)}{\Lambda(E)}\right]\right\} d\theta \cdot \int_0^L exp\left(-\frac{x}{\Lambda_{out}(E_{out})}\right) dx \quad (1)$$

The first factor represents the energy distribution of the thermal-neutron flux Φ from the premoderator at temperature $T$, incident on the p-$H_2$ moderator. It is described by a Maxwellian energy distribution, where $k_B = 0.08617$ meV/K is Boltzmann's constant:

$$M(E,T) = \frac{E}{(k_B T)^2}\, exp\left(-\frac{E}{k_B T}\right) \quad (2)$$

The second factor describes the production of cold neutrons through para-to-ortho conversion of a "parent" thermal neutron with energy $E$ and mean free path $\Lambda(E)$ in p-$H_2$ into a cold neutron with energy $E_{out} = E - \Delta E$ , within the single-collision approximation. The factor $cos(\theta)$ accounts for the Lambertian angular distribution of the thermal-neutron flux Φ emitted by the premoderator as a result of the thermalization of fast neutrons produced in the target. The function $S(\theta, H, L)$ describes the thermal-neutron path lengths, averaged over the entire length of the p-$H_2$ moderator, for neutrons emitted from different locations on the premoderator (see Appendix 1).

The third factor describes the accumulation of cold neutrons with a mean free path $\Lambda_{out}(E_{out})$ in the p-$H_2$ moderator of length $L$ (Fig. 2). Assuming uniform illumination of the moderator, this accumulation is described by the following function:

$$J(L, E_{out}) \sim \int_0^L exp\left(-\frac{x}{\Lambda_{out}(E_{out})}\right) dx = \Lambda_{out}(E_{out}) \left[1 - exp\left(-\frac{L}{\Lambda_{out}(E_{out})}\right)\right] \quad (3)$$

Figure 3 shows the dependence of the emerging-beam intensity on moderator length for low-energy neutrons with an energy of 9 meV (wavelength $\lambda$ = 3Å), whose mean free path in p-$H_2$ is approximately 19 cm, and for higher-energy neutrons with energies of 20 meV ($\lambda$ = 2Å), 25 meV ($\lambda$ =1.8 Å) and 82 meV ($\lambda$ =1 Å), whose mean free paths in p-$H_2$ are less than 3.7 cm (see Fig. 1). Since the mean free paths of neutrons with energies of 2–9 meV (corresponding to wavelengths of approximately 2.6–6.4 Å) are nearly identical and amount to about 19 cm (Fig. 1), the curve shown for 9 meV neutrons is also representative of this entire energy range.

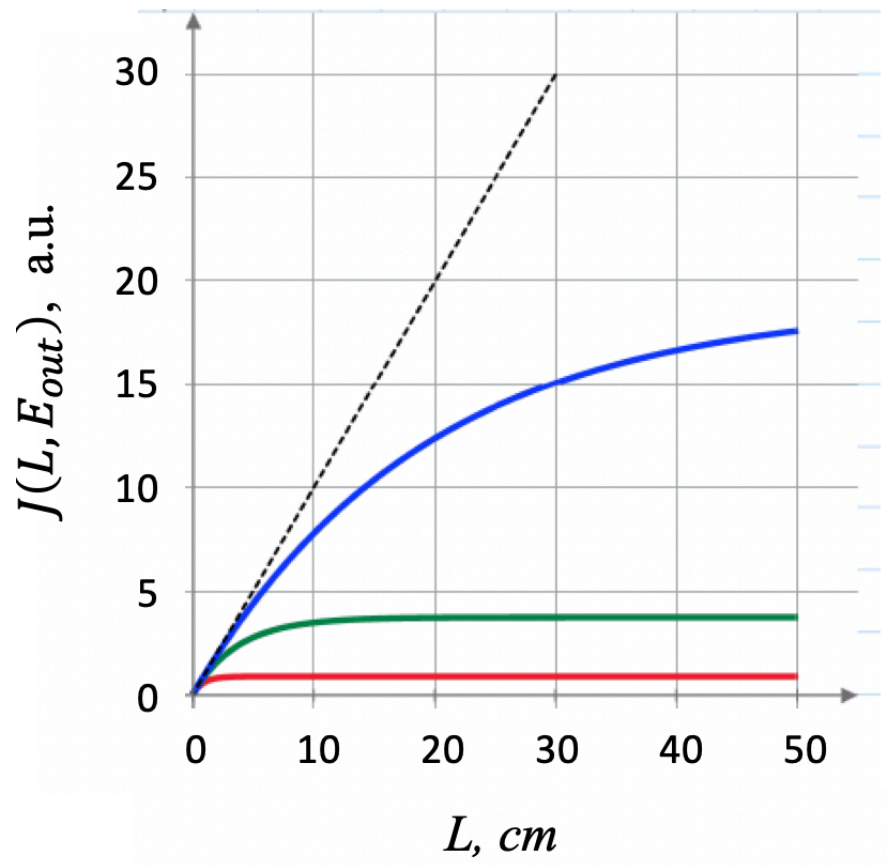

**Fig. 3.** Dependence of the emerging neutron-beam intensity on moderator length for neutrons with energies of 82 meV (red line), 20 meV (green line) and 9 meV (blue line). The dashed line represents the dependence for the idealized case of an infinite mean free path.

Figure 3 shows that, although the mean free path of cold neutrons in p-$H_2$ is relatively long (approximately 19 cm), its finite value limits the accumulation of neutron intensity. The effective moderator length $L_{eff}$ over which this accumulation remains significant is approximately three mean free paths of the emerging neutrons.

For example, for neutrons with an energy of 3 meV ($\Lambda_{out}$ ≈ 19 см), the intensity at $L_{eff}$ = 50 cm reaches approximately 94% of its asymptotic value. For neutrons with an energy of 20 meV ($\Lambda_{out}$ ≈ 3.7 см), saturation is reached at $L_{eff}$ = 11 cm, while for neutrons with an energy of 82 meV ($\Lambda_{out}$ ≈ 1 см), it occurs at $L_{eff}$ = 3 cm. Thus, increasing the moderator length beyond these values provides virtually no benefit for the corresponding neutron energies, since only neutrons generated near the end of the moderator can escape and contribute to the emerging beam.

It should be noted that the single-collision approximation used in Eq. (1) is valid only for thin moderators containing a relatively small volume of p-$H_2$ [8, Fig. 24b]. It does not apply to thick moderators containing a larger p-$H_2$ volume, where the probability of a second collision can be significant, reaching 50% [8, Fig. 23b].

## 2.2. Accelerator-power utilization efficiency

In the ideal case of an infinite mean free path, the intensity of the neutron beam emerging from the moderator is proportional to the moderator length ***L*** (dashed line in Fig. 3). As discussed in Section 2.1, increasing the moderator length requires a proportional increase in the proton-beam intensity and, consequently, in the accelerator power ***P***.

For HiCANS-class sources, a maximum allowable target heat flux of approximately 0.5–1 kW/cm² is generally accepted, depending on the target material [11]. A representative example is the HBS project, which employs a tantalum target with microchannel cooling [7], one of the most effective technical solutions currently available. For a target area of (10 x10) cm², the maximum proton-beam power is 0.1 MW. Taking this value as a reference and assuming that the moderator width $W$ remains constant at 10 cm throughout this work, we obtain the following relationship between the moderator length $L$ and the required total incident proton-beam power $P$:

$$L = \beta \cdot P$$

where $\beta$ = 100 cm/MW.

The efficiency of accelerator-power utilization as the moderator length increases can then be defined as the ratio of the actual intensity emitted by a moderator of length $L$ to the intensity emitted by an ideal moderator of the same length (here ideal means fully transparent for the outgoing neutrons of the given wavelength):

$$\eta(L, E_{out}) = \frac{I_{out}(E_{out}, L)}{I_{ideal}(E_{out}, L)} = \frac{J(L, E_{out})}{L} \sim \frac{\Lambda_{out}(E_{out}) \cdot \left[1 - exp\left(-\frac{L}{\Lambda_{out}(E_{out})}\right)\right]}{L} \quad (4)$$

Alternatively, this efficiency can be expressed as a function of accelerator power $P$:

$$\eta(P, E_{out}) \sim \frac{\Lambda_{out}((E_{out}) \cdot \left[1 - exp\left(-\frac{\beta \cdot P}{\Lambda_{out}(E_{out})}\right)\right]}{\beta \cdot P} \quad (5)$$

The calculated dependence of the accelerator-power utilization efficiency $\eta$ as a function of accelerator power $P$ for different neutron energies are presented in Fig. 4. It can be seen that simply increasing the p-$H_2$ moderator length while proportionally increasing the accelerator power is inefficient.

For 3-meV neutrons ($\lambda$ =5 Å), at an accelerator power of $P$ = 0.1 MW ($L$ = 10 cm, $W$ = 10 cm), the accelerator-power utilization efficiency is $\eta \approx 78\%$. However, a fivefold increase in accelerator power to $P$ = 0.5 MW ($L$ =50 cm) increases the intensity of the emerging neutron beam by only a factor of about 2.3 (see Fig. 3), while the accelerator-power utilization efficiency decreases to 35%. This suggests that, in principle, the useful neutron intensity can be increased by a factor of 2–2.5 at the same accelerator power.

For higher-energy neutrons, for example, 20-meV neutrons ($\lambda$ =2 Å), which have shorter mean free paths, the accelerator-power utilization efficiency at $P$ = 0.1 MW is approximately half that for 3-meV neutrons, $\eta \approx 34\%$ (Fig. 4). Meanwhile, the emerging-beam intensity is already close to saturation, exceeding 90% of its maximum attainable value (Fig. 3). Therefore, increasing the proton-beam power to 0.5 MW would not lead to any significant increase in the emerging-beam intensity, but would reduce $\eta$ to approximately 7%.

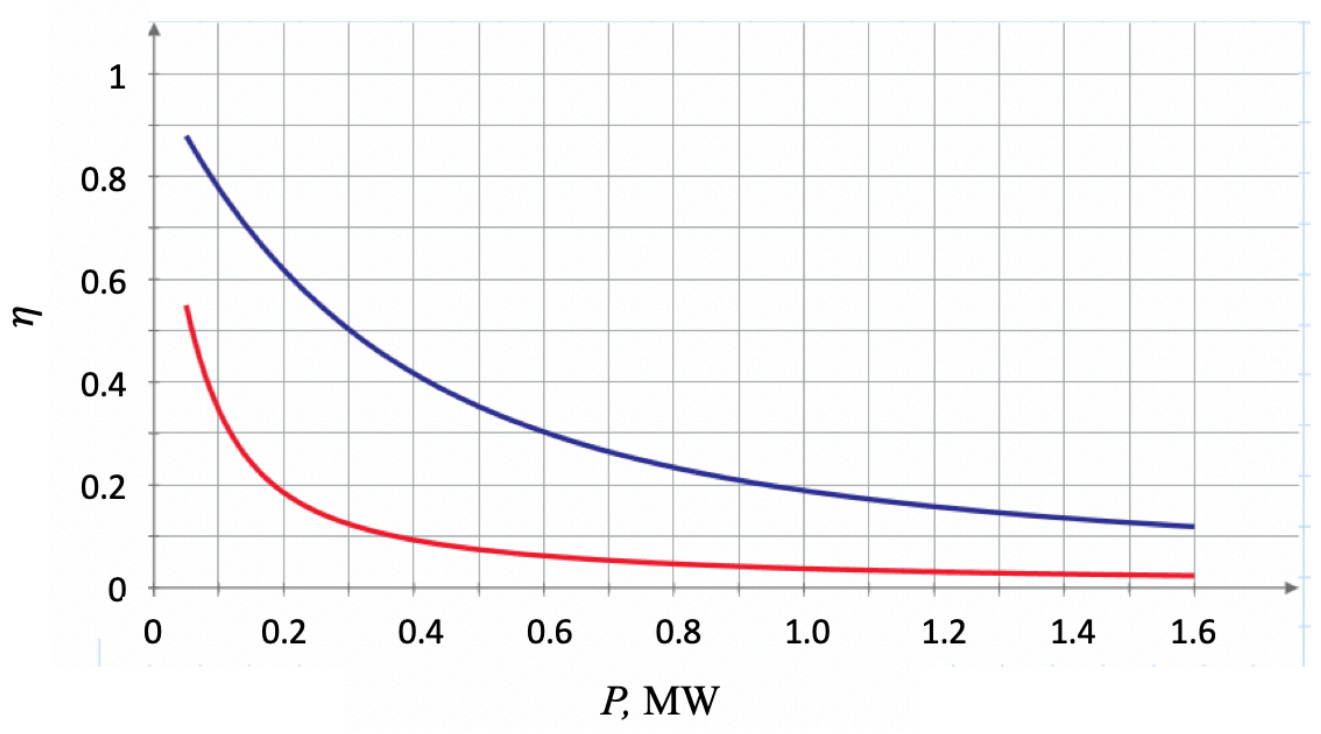

Fig. 4. Accelerator-power utilization efficiency as a function of moderator length for neutron wavelengths of 20 meV ($\lambda$ = 2Å) (red) and 3 meV ($\lambda$ =5 Å) (blue).

Thus, a proportional increase in accelerator power $P$ with increasing total moderator length $L$ does not provide an effective means of increasing the intensity of the emerging neutron beams. The maximum achievable cold-neutron flux from a p-$H_2$ moderator is limited primarily not by the maximum available accelerator power, but by the permissible heat load on the target, which determines its dimensions. This points to the need for new geometrical approaches to the target–premoderator system.

It should be emphasized that heat-removal limitations are not confined to compact neutron sources. Similar constraints arise in spallation neutron sources with stationary targets, such as ISIS and CSNS, where, unlike at ESS, the proton beam is concentrated on a limited region of the target rather than distributed over the surface of a rotating disk. In this case, too, further increases in accelerator power are limited by the maximum permissible heat load on the target.

## 3. Staircase moderators with adapted target-premoderator units.

One possible way to improve the utilization of increased accelerator power is to employ the staircase assembly of low-dimensional (compressed) p-$H_2$ moderators proposed earlier [8]. This arrangement retains the high brightness of individual compressed moderators while restoring a large transverse cross-section of the emerging cold-neutron beam. Under comparable angular-acceptance conditions, the increased overall beam aperture can compensate for the intensity loss associated with an individual compressed moderator, allowing the intensity gain of the combined assembly to approach its brightness gain.

However, as demonstrated in [12], the practical gain achieved with the staircase assembly may be substantially smaller than its potential maximum. This is due both to the limited extent of the thermal-neutron generation region along the premoderator and to the varying distances of the individual steps from the premoderator. These factors lead to non-uniform illumination of the steps in the staircase moderator assembly and, consequently, significantly reduce its overall effectiveness.

To realize the full potential gains in brightness and intensity, the target–premoderator geometry must therefore be adapted to the staircase moderator assembly to ensure efficient thermal-neutron illumination of each step. The most natural solution is a staircase arrangement of target–premoderator units that follows the geometry of the staircase p-$H_2$ cold source (Fig. 5b). Each target–premoderator unit then supplies thermal neutrons to its corresponding moderator step, thereby creating a more extended region of uniform illumination while retaining the benefits of compressed moderators. The relatively short length of each step ensures efficient utilization of the proton-beam power.

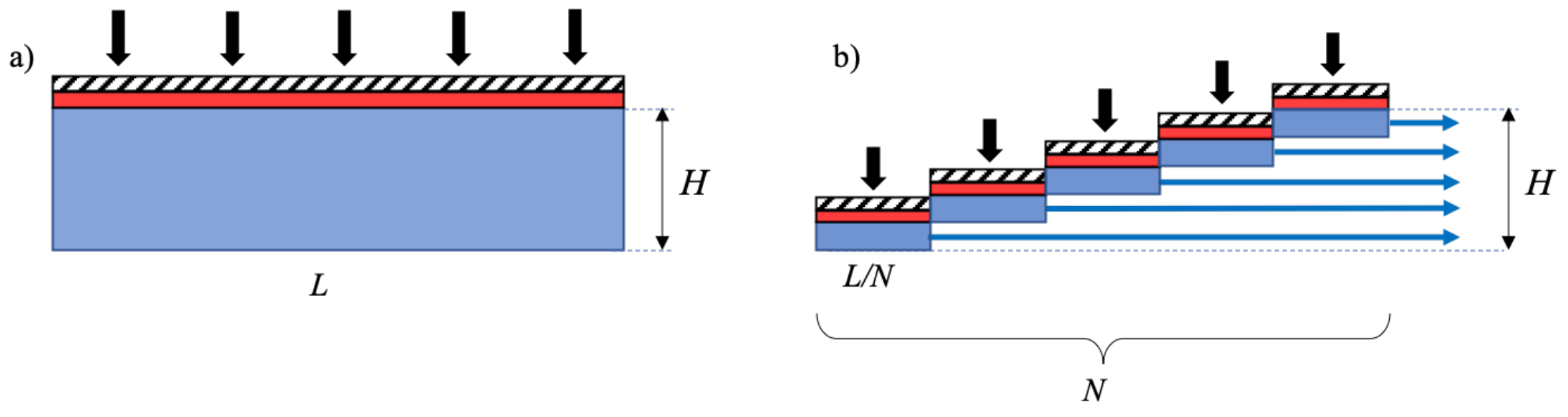

Fig. 5. (a) Single target–premoderator–moderator assembly; (b) staircase assembly comprising $N$ steps, each with its own target–premoderator unit and p-$H_2$ moderator. The width of the assemblies, measured perpendicular to the plane of the figure, is $W$ (see Fig. 2).

### 3.1 Brightness and intensity of neutron beams from a staircase assembly

We compare the neutron characteristics of a staircase assembly consisting of $N$ identical units, each comprising a target–premoderator (TP) and a p-$H_2$ moderator of the same length (Fig. 5b), with those of the reference configuration, consisting of a single TP unit and a p-$H_2$ moderator, corresponding to a single step (($N = 1$); Fig. 5a). The voluminous moderator is chosen as the reference because the total target area is the same in both configurations and, consequently, the accelerator power used is also the same. Both configurations have the same overall external moderator dimensions ($W \times L \times H$), so the staircase assembly fits within the envelope of the reference moderator.

It should be noted that the particular choice of a moderator height of 10 cm is not essential here. The results would not differ significantly for another specified height, for example, 3 cm. Some practical limitations associated with the choice of this dimension will be discussed in Section 4.

We define the brightness of a moderator $B_{out}(E_{out}, L, H, T)$ as the ratio of the neutron intensity $I_{out}(E_{out}, L, H, T)$ emitted normal to its surface ($W \times H$) to the area $W \cdot H$ of that surface. Here, we assume that the distance to the detector measuring the neutron intensity is fixed and much larger than the moderator dimensions:

$$B_{out}(E_{out}, L, H, T) = \frac{I_{out}(E_{out}, L, H, T)}{W \cdot H}$$

The brightness gain of an individual step relative to the reference moderator is given by the product of the gain $G_H(E_{out}, N, L, H, T)$ due to compressing the moderator by a factor of $N$ to a step height $h = H/N$ and the reduction factor $G_L(E_{out}, N, L)$ associated with shortening the moderator by a factor of $N$, to a step length $L/N$ (Fig. 5b):

$$G_B^{Step}(E_{out}, N, L, H, T) = G_H(E_{out}, N, L, H, T) \cdot G_L(E_{out}, N, L) \qquad \text{[Eq. (6)]}$$

Here

$$G_H(E_{out}, N, L, H, T) = \frac{B_{out}(E_{out}, L, H/N, T)}{B_{out}(E_{out}, L, H, T)}$$

[Eq. (7)]

where $B_{out}(E_{out}, L, \frac{H}{N}, T)$ and $B_{out}(E_{out}, L, H, T)$ are the brightnesses of the compressed and reference para-hydrogen moderators, respectively.

Figure 5 shows the brightness gains $G_H(E_{out}, N, L, H, T)$ calculated using PHITS [13] within the single-collision approximation, as a function of moderator height $h = H_0/N$, relative to the brightness of voluminous moderator with a height of $H_0$ =10 cm. The moderator is assumed to be uniformly illuminated by a thermal-neutron flux with a Maxwellian energy distribution and a Lambertian angular distribution (see Eq. (1)). The energy dependence of the brightness gain arises from the energy dependence of the mean free path of incident neutrons in p-$H_2$ (Fig. 1).

The quantity

$$G_L(E_{out}, N, L) = \frac{J(E_{out}, L/N)}{J(E_{out}, L)}$$

**[Eq. (8)]**

describes the reduction in brightness associated with shortening an individual moderator step. Figure 6 shows the dependence of this reduction factor on the number of steps $N$ for a moderator with a total length of $L$ =50 cm. The curves for neutrons with energies of 3, 5, and 9 meV almost completely overlap. This is because the mean free paths of cold neutrons in this energy range are nearly identical, amounting to approximately 19–20 cm (Fig. 1).

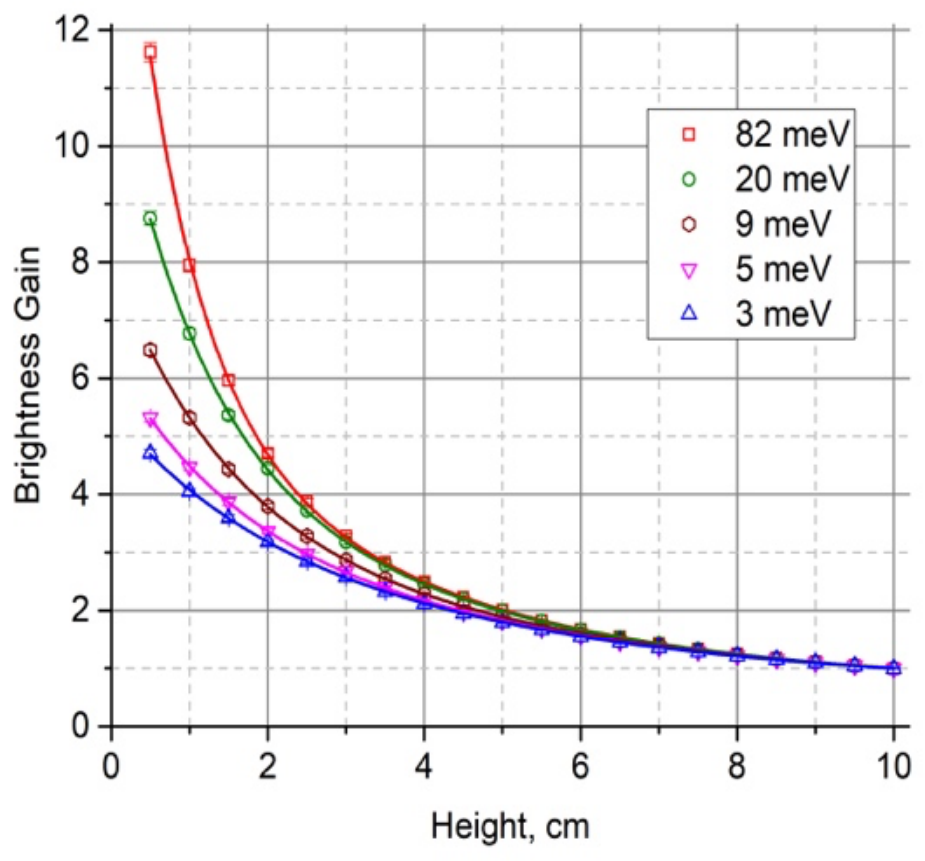

**Fig. 6.** Brightness gain as a function of moderator height $h$, normalized to the brightness of a moderator with a height of $H_0$=10 cm, for emerging neutrons with energies of 82, 20, 9, 5, and 3 meV, corresponding to wavelengths of 1, 2, 3, 4, and 5 Å, respectively.

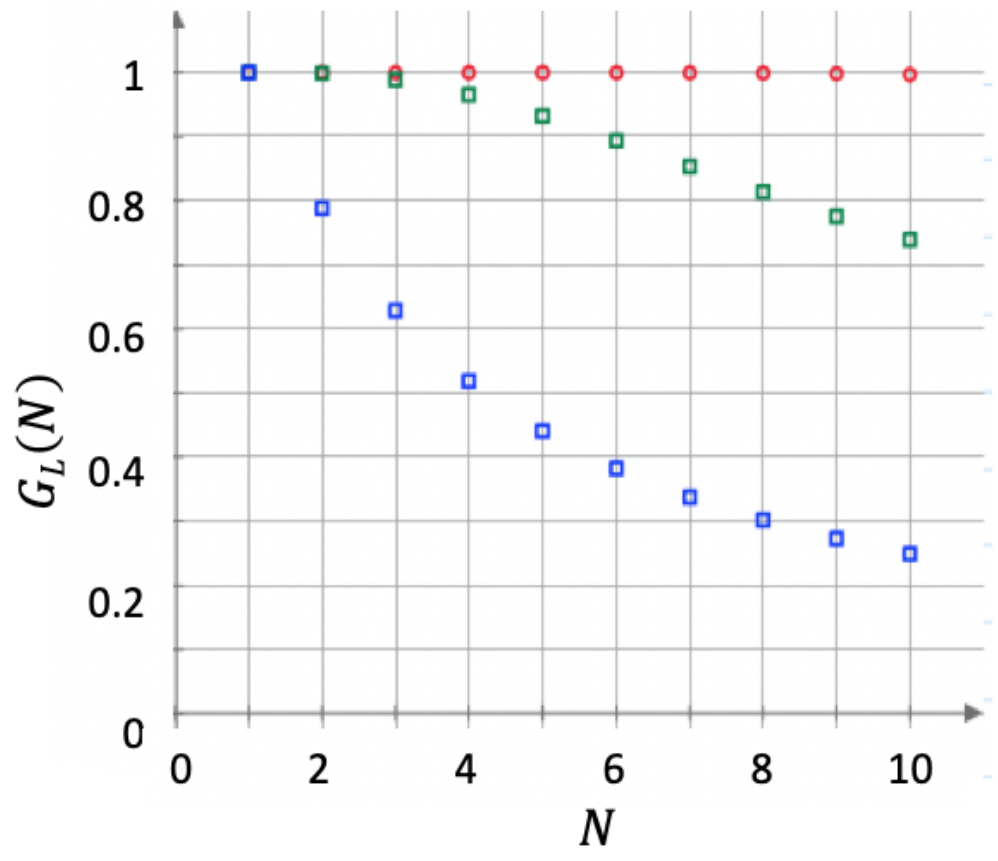

**Fig. 7.** Brightness reduction factor $G_L(N, L, E_{out})$ as a function of the number of steps $N$ for a moderator with a total length of $L$ =50 cm, for emerging neutrons with energies of 82, 20, 9, 5, and 3 meV, corresponding to wavelengths of 1, 2, 3, 4, and 5 Å, respectively.

Since the intensity is proportional to the product of the moderator brightness and its emitting surface area, the intensity emitted by a single step can be expressed as

$$I_{Step}(E_{out}, N, L, H, T) \equiv I_{out}\left(E_{out}, \frac{L}{N}, \frac{H}{N}, T\right) =$$

$$= B_{out}\left(E_{out}, \frac{L}{N}, \frac{H}{N}, T\right) \cdot \frac{H}{N} \cdot W \equiv B_{step}(E_{out}, N, L, H, T) \cdot \frac{H}{N} \cdot W$$

**[Eq. (9)]**

The total intensity from the staircase assembly is obtained by summing the intensities of the beams emerging from the individual steps. Accounting for the reduction factor $F_{wall}$ associated with the structural walls of thickness $a$ separating adjacent steps (see Section 3.2), we obtain:

$$I_{Stair}(E_{out}, N, L, H, T, a) = I_{Step}(E_{out}, N, L, H, T) \cdot N \cdot F_{wall}(E_{out}, N, H, a)$$

**[Eq. (10)]**

For $N$=1, the loss factor $F_{wall}$ equals unity, and the intensities $I_{Stair}$ and $I_{Step}$ coincide:

$$I_{Stair}(E_{out}, 1, L, H, T) = I_{Step}(E_{out}, 1, L, H, T).$$

The brightness of the staircase assembly is given by

$$B_{Stair}(E_{out}, N, L, H, T, a) = \frac{I_{Stair}(E_{out}, N, L, H, T, a)}{H \cdot W}$$

Using Eq. (10), we obtain

$$B_{Stair}(E_{out}, N, L, H, T, a) = \frac{I_{Step}(E_{out}, N, L, H, T)}{\frac{H}{N} \cdot W} \cdot F_{wall}(E_{out}, N, H, a)$$
$$= B_{step}(E_{out}, N, L, H, T) \cdot F_{wall}(E_{out}, N, H, a)$$

Thus, the brightness of the staircase assembly equals that of an individual compressed step, reduced by the loss factor associated with the material walls surrounding the p-$H_2$.

At the same time, the staircase assembly provides a substantial gain in intensity $G_I^{Stair}$ relative to the reference moderator, for which $N$ =1 (Fig. 5a). Using Eq. (9), we obtain

$$G_I^{Stair}(E_{out}, N, L, H, T, a) = \frac{I_{Stair}(E_{out}, N, L, H, T, a)}{I_{Stair}(E_{out}, 1, L, H, T, a)}$$
$$= \frac{I_{Step}(E_{out}, N, L, H, T) \cdot N}{I_{Step}(E_{out}, 1, L, H, T)} \cdot F_{wall}(E_{out}, N, H, a)$$
$$= \frac{B_{step}(E_{out}, N, L, H, T) \cdot \frac{H}{N} \cdot W}{B_{step}(E_{out}, 1, L, H, T) \cdot H \cdot W} \cdot N \cdot F_{wall}(E_{out}, N, H, a)$$

Finally,

| $G_I^{Stair}(E_{out}, N, L, H, T, a) = G_B^{Step}(E_{out}, N, L, H, T) \cdot F_{wall}(E_{out}, N, H, a)$ | (11) |
|---|---|

Thus, the intensity gain provided by a staircase assembly comprising $N$ moderators with a total length of $L$ and an aperture of $(W \times H)$, relative to a reference moderator of dimensions $(W \times H \times L\ )$, is equal to the brightness gain resulting from the compression and shortening of the individual moderator steps by a factor of $N$ . This differs fundamentally from compressing the reference moderator by a factor of $N$ to dimensions of $(W \times H/N \times L)$, a process in which the increase in the brightness of the emitted beam is accompanied by a decrease in its intensity.

It should be emphasized that the volume of para-hydrogen used in the staircase assembly, and consequently the heat deposited in it, is $N$ times smaller than in the reference moderator.

### 3.2 Brightness Losses Due to Structural Walls

The structural walls forming the moderator vessel partially shadow the emitting surfaces of the para-hydrogen steps in the staircase assembly. In Ref. [8], a conservative case of perfectly neutron-opaque walls was considered. In this case, the staircase geometry contains non-emitting layers (see Fig. 8a), and the loss factor in Eq. (10) can be expressed as follows:

$$F_{wall}(E_{out}, N, H, a) = 1 - a\frac{N-1}{H}$$

**[Eq. (12)]**

This model provides an upper-bound estimate of the losses caused by the finite thickness of the para-hydrogen moderator vessel walls.

In practice, cryogenic moderator vessels are made of Al6061 alloy, which exhibits appreciable transparency to cold neutrons. This property allows the overlap geometry of adjacent steps to be modified. Unlike the configuration shown in Fig. 8a, the aluminum walls can partially overlap the emerging neutron beams from adjacent steps (Fig. 8b), since some neutrons pass through the vessel wall material with only minor attenuation.

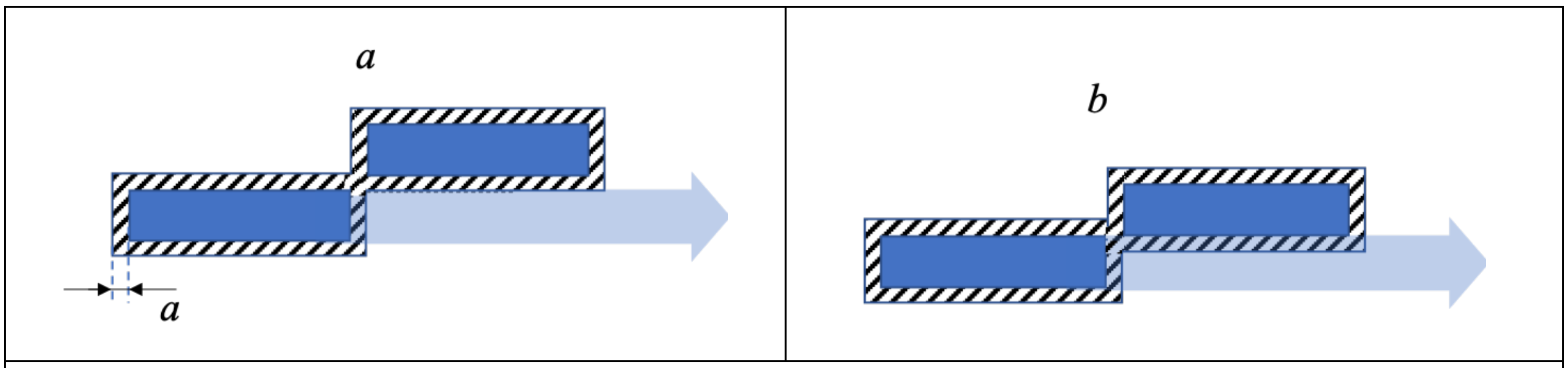

**Fig. 8.** Overlap geometry of adjacent moderator steps separated by walls of thickness (d): (a) horizontal walls that are completely opaque to neutrons; (b) horizontal walls that are partially transparent to neutrons.

In this case, the loss factor should be expressed as

$$F_{wall}(E_{out}, N, L, H, a) = 1 - a\frac{N-1}{H}\left(1 - \exp\left(-\sigma_{wall}(E) \cdot n_{wall} \cdot \left(\frac{L}{N} + a\right)\right)\right)$$

**[Eq. (13)]**

where the total cross section $\sigma_{wall}(E)$ and the atomic number density $n_{wall}$ of the wall material determine the attenuation of the emerging neutron beam.

Using the cross-section values obtained in Ref. [14], the attenuation of a neutron beam passing through a layer of Al6061 alloy cooled to 20 K ($n_{wall}$= $6.25\times10^{22}$ атом/cm$^3$) can be calculated. For practical values of $L$=50 cm, $N$=5 and $a$ = 0.2 cm, the transmission factors are approximately 0.49 and 0.93 for neutrons with energies of 3.3 meV ($\lambda$ =5 Å) and above 20 meV ($\lambda$ < 2 Å), respectively. Equation (46) then gives loss factors of approximately 0.96 and 0.99, respectively (Fig. 9). Thus, the brightness gains are reduced by only 10–15% compared with the configuration with completely opaque walls (Eq. (12)), and the resulting losses in intensity amount to only a few percent.

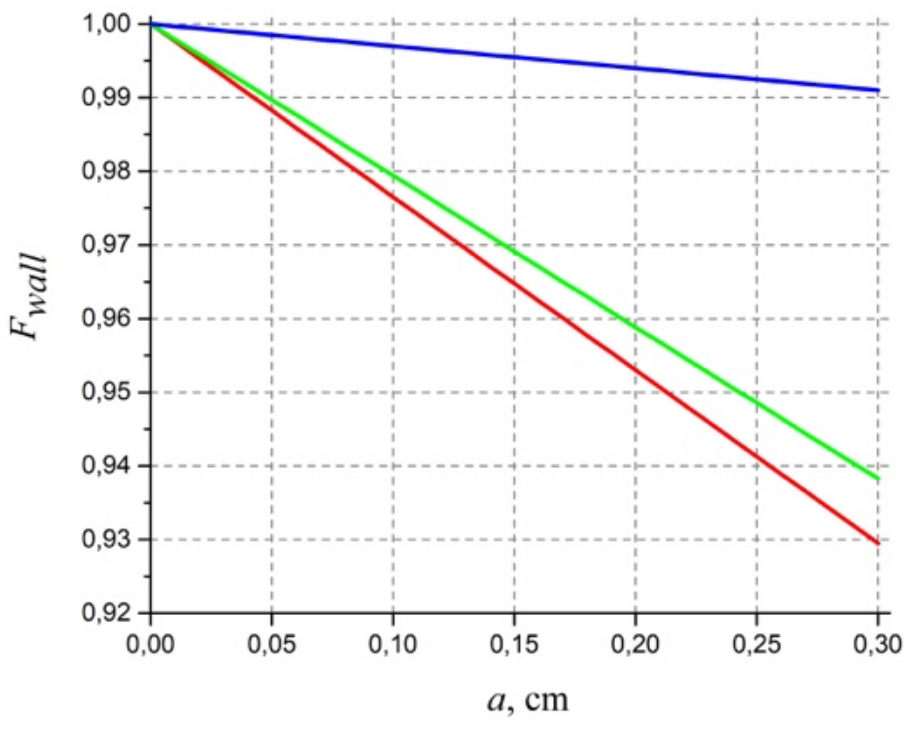

Fig. 9. Loss factor of a five-step staircase assembly as a function of wall thickness $a$, accounting for neutron transmission through Al6061 alloy at 20 K. The curves correspond to neutron energies of 82 meV (red), 20 meV (green), and 3.3 meV (blue).

### 3.3 Comparison with PHITS Simulations

Top panel of Fig. 10 presents the brightness spectra obtained from PHITS simulations [13] for the staircase assembly (configuration shown in Fig. 5b), consisting of five steps, each 10 cm long, and for the reference moderator with dimensions of $(10 \times 10 \times 50)$ cm³. In both cases, the moderators were irradiated by a proton beam with a total power of $P$ = 0.5 MW.

Bottom panel of Fig. 10 shows the brightness gain, calculated as the ratio of the brightness of the staircase assembly to that of the reference moderator. The figure also shows the analytically calculated gains obtained using Eq. (11) for neutron energies of 3, 5, 9, 20 and 82 meV, corresponding to wavelengths 5, 4, 3, 2 и 1 Å, respectively (red circles). The same moderator parameters were used in the analytical calculations.

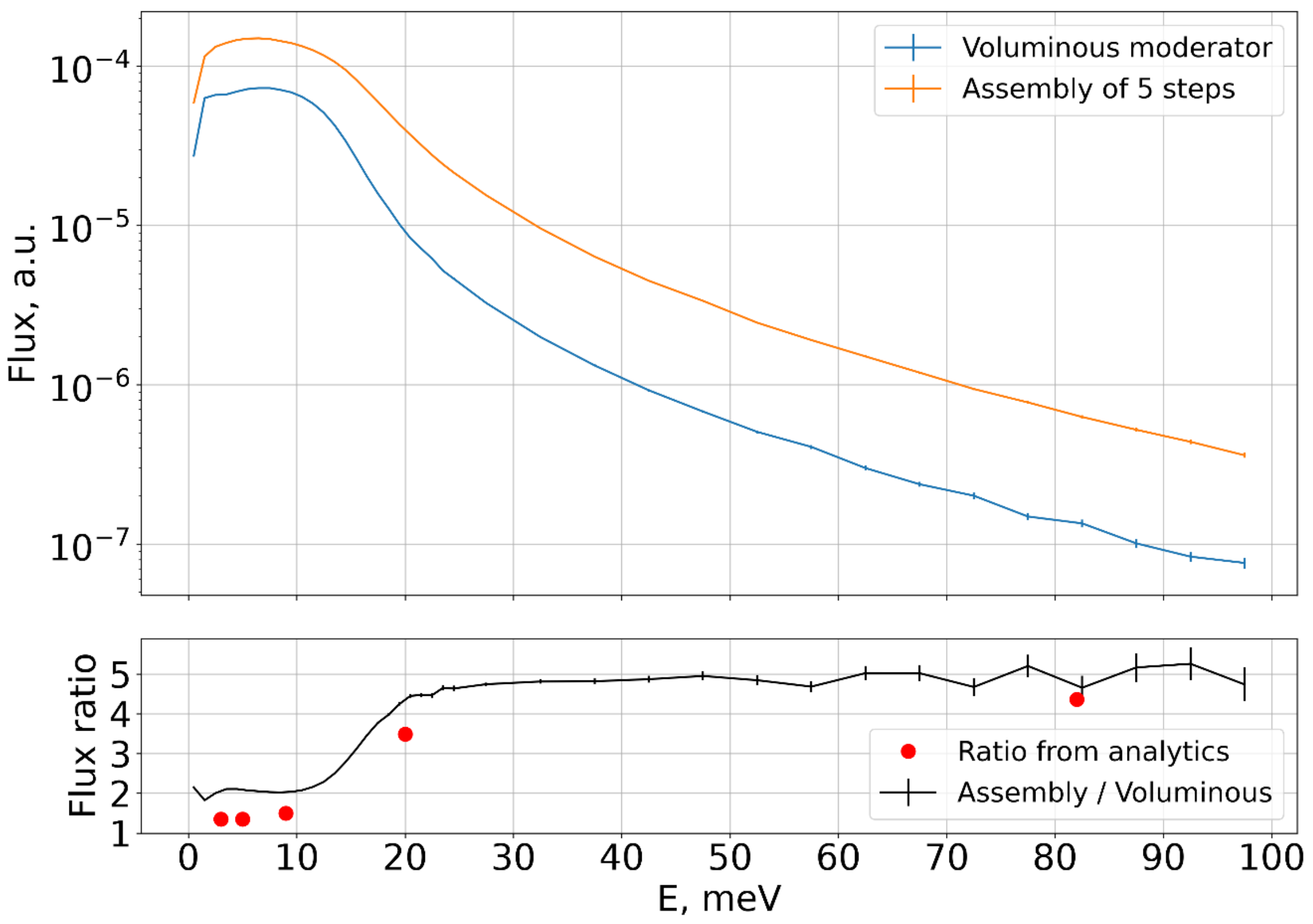

Fig. 10. Top: spectra of the 5-step staircase moderator (orange) and of the voluminous moderator (blue), calculated with PHITS within single-collision approximation. Bottom: the ratio of these spectra (black line) and results obtained analytically for the energies of 3, 5, 9, 20 and 82 meV (red dots).

Good agreement is observed between the PHITS simulation results and the analytical calculations: the dependences have similar shapes, and the difference in intensity is approximately 10–15%. This difference may be attributed to additional illumination of the end faces of the para-hydrogen moderator steps by thermal neutrons arriving from adjacent water premoderators. This effect is not taken into account in the analytical model.

Thus, the PHITS simulation results support the validity of the analytical model, which will be used for further analysis.

## 4. Optimization of the Staircase Assembly

### 4.1. Unlimited total accelerator power

Let us assume that the total accelerator power $P$ is unlimited. Since $P{\sim}L$ (see Section 2.1), this is equivalent to assuming that there is no upper limit on the total length $L$ of the system.

As shown by Eq. (11), the gain provided by the staircase cold source are determined primarily by the product of two factors, $G_B(H/N, E_{out})$ and $G_L(N, L, E_{out})$, which have different dependences on the number of steps $N$. Here we neglect the wall-loss factor, which depends only weakly on $N$ (see Sect. 3.2). Increasing the number of steps while keeping the total moderator height $H$ fixed reduces the height $h = H/N$ of each step and, consequently, increases the brightness of individual steps due to the compression effect (Fig. 6). At the same time, however, the length of each step $L/N$ decreases, reducing the intensity of the emerging beam because less cold-neutron intensity accumulates along each step (Fig. 7). The competition between these two factors determines the existence of an optimum number of steps and an optimum moderator length, and hence an optimum accelerator power $P$, at which the gain provided by the staircase assembly reaches its maximum.

The calculated brightness and intensity gains for a staircase assembly with a total height of $H$ = 10 cm, relative to a single moderator with the same external dimensions, are shown in Fig. 11 for neutrons with energies of 3 meV ($\lambda$ =5 Å), 20 meV ($\lambda$ =2 Å) and 82 meV ($\lambda$ =1 Å). As expected, the dependences on $P$ and $N$ exhibit pronounced maxima. The gains reach relatively high values of approximately 3, 5, and 10, respectively.

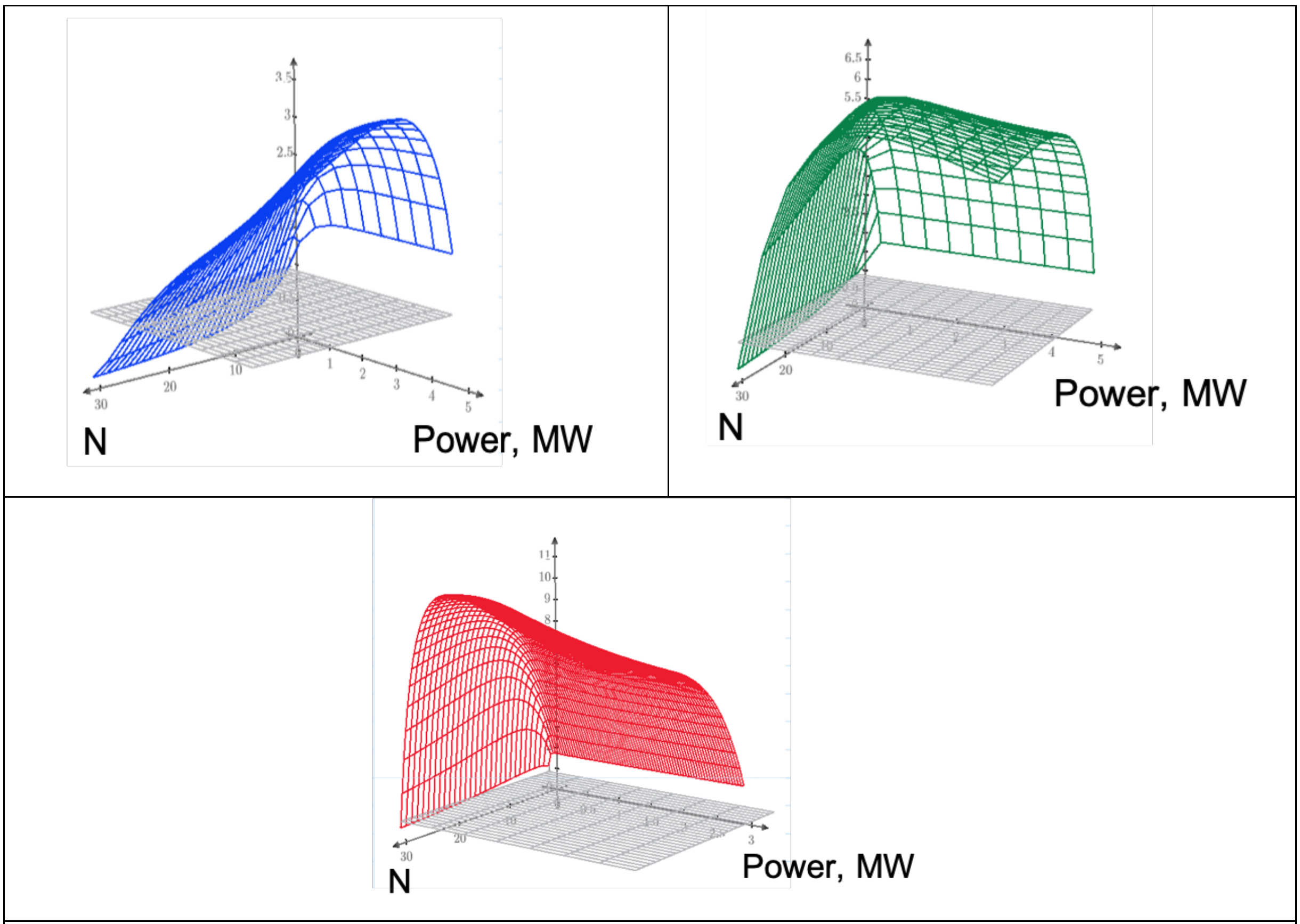

Fig. 11. Three-dimensional plots of the intensity gain provided by the staircase assembly as a function of its total length $L$ and the number of steps $N$ relative to a single moderator with the same external dimensions ($N$=1): (a) for neutrons with an energy of 3 meV; (b) for neutrons with an energy of 20 meV; (c) for neutrons with an energy of 82 meV. The grey plane indicates unity gain.

Thus, the proposed staircase target–moderator assembly, in which each moderator step is illuminated by its own target unit of an appropriate size operating at the same maximum allowable heat flux, effectively removes the limitation on usable accelerator power imposed by the finite dimensions of a single target. This makes it possible to employ longer moderators and, consequently, higher-power proton beams.

Fig. 12 presents cross-sections of the 3D surfaces shown in Fig. 11, illustrating the dependence of the gains on the total accelerator power $P$ for a staircase assembly with different numbers of steps $N$. The brightness and intensity gains are nearly identical (see Eq. (11)), as the wall factors are close to unity (see Sect. 3.2).

For 3-meV neutrons (Fig. 12a), the gain increases almost linearly with accelerator power up to $P \approx$ (2–2.5) MW, reaching approximately 3.5. At this power, the accelerator-power utilization efficiency $\eta$ (see Eq. (4)) is close to 100%. Beyond this range, the gain rapidly approaches saturation, while $\eta$ decreases approximately as $1/P$.

For 20-meV neutrons (Fig. 12b) with shorter mean free path, the linear-growth region extends only to $P \approx$ (0.5–1) MW, where the gain reaches approximately 5.5 and $\eta$ remains close to 100%. At higher powers, the gain likewise approaches saturation, with a corresponding decrease in $\eta$ .

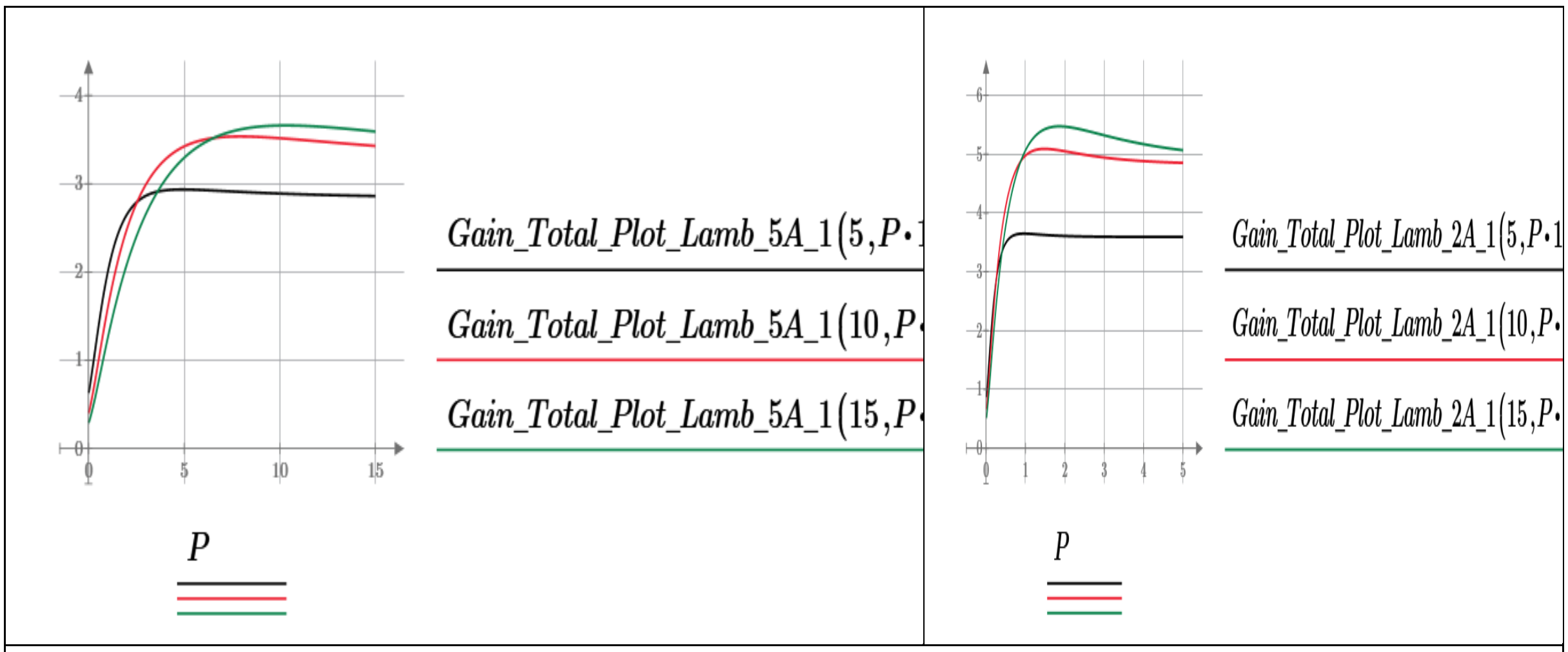

Fig. **12**. Brightness and intensity gains provided by the staircase assembly as functions of the total proton-beam power $P$ for different numbers of steps $N$: (a) for neutrons with an energy of 3 meV; (b) for neutrons with an energy of 20 meV. The black, red, and green curves correspond to $N$ = 5, $N$ = 10, and $N$ = 15, respectively.

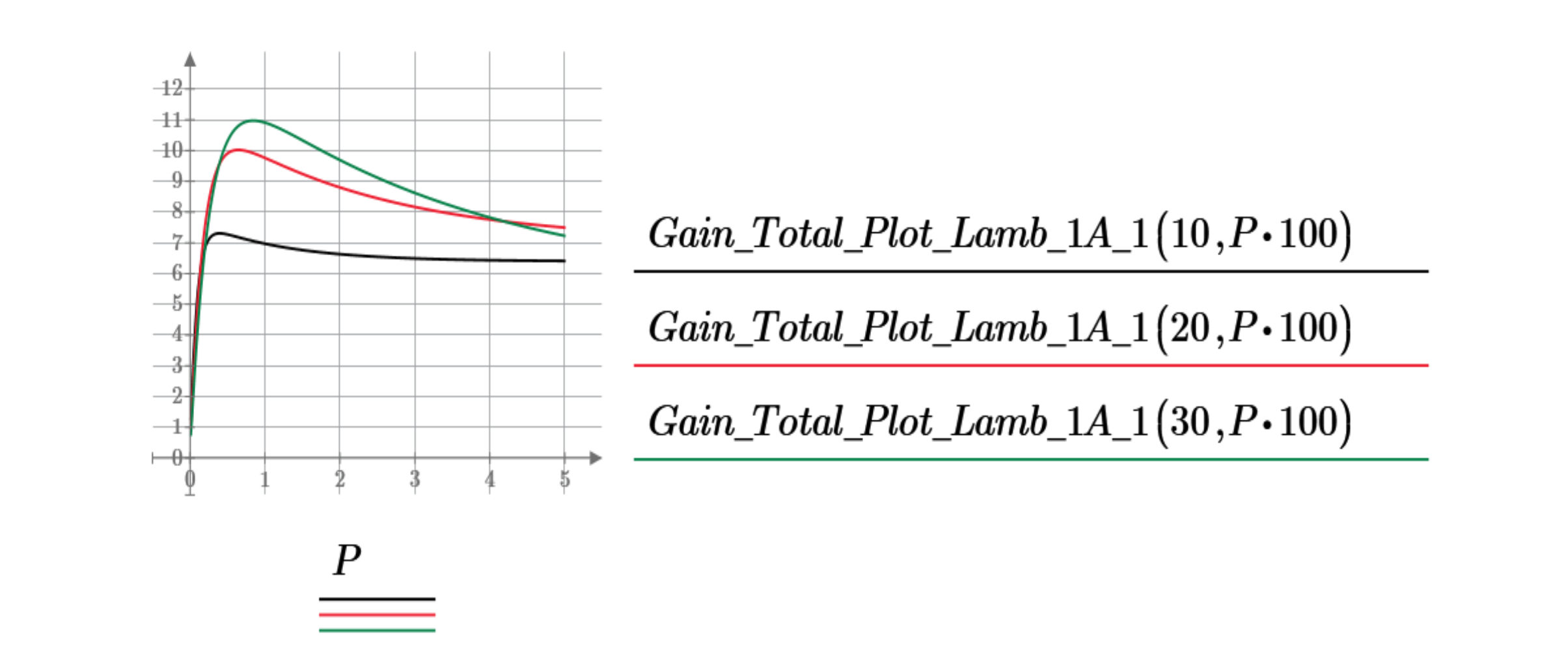

Fig. 13. Brightness and intensity gains provided by the staircase assembly as functions of the total proton-beam power $P$ for different numbers of steps $N$: (a) for neutrons with an energy of 82 meV. The black, red, and green curves correspond to $N$ = 10, $N$ = 20, and $N$ = 30, respectively.

Consequently, using an accelerator with a power exceeding approximately 2.5 MW is inefficient for the configuration considered here. This limitation is determined by the total height of the staircase assembly, $H$ =10 cm: since the accelerator power $P$ is proportional to the total moderator length $L$ , increasing $L$ requires an increase in the number of steps $N$. At large $N$, the height $h$ of each step becomes comparable to the wall thickness, and the wall-loss factor (see Eq. (12)) suppresses the gain [8,12]. Note, that increasing the total moderator height $H$ shifts the upper power limit for efficient operation proportionally towards higher accelerator powers.

The substantially greater intensity gain for 20-meV neutrons is explained by their short mean free path in p-$H_2$, $\Lambda_{out} \approx 3.7$ cm, compared with about 19 cm for 3-meV neutrons. This results in low transparency of p-$H_2$ to 20-meV neutrons: approximately 95% of these neutrons are attenuated over a moderator length of about $3\Lambda_{out}$. Consequently, only the final section of the moderator, approximately 11 cm long, effectively contributes to the emerging beam. The contribution of the upstream portion of a moderator longer than 11 cm to the emerging-beam intensity is therefore almost entirely eliminated. Dividing a long moderator into shorter steps of approximately 11 cm increases the efficiency of producing and extracting 20-meV neutrons from each step. This behavior is evident in Fig. 12b, where the onset of saturation for the staircase assembly occurs at an accelerator power of approximately 0.5 MW ($L$ =50 cm) for $N$ =5, at 1 MW ($L$ =100 cm) for $N$ =10 and at 1.5 MW ($L$ =150 cm) for $N$ =15: in all cases the step length is about 10cm.

For neutrons with energies above 82 meV, which have even shorter mean free paths, the optimal step length is correspondingly smaller. For example, for 82-meV neutrons with a mean free path of approximately 1 cm, the optimal step length is about 3 cm, and the intensity gain increases to approximately 10–11 (Fig. 13). This gain is limited by the loss factor associated with the aluminum moderator walls, since subdividing the moderator into more steps also reduces the height of each step, eventually making it comparable to the wall thickness [8].

## 4.2 Limited total accelerator power

### 4.2.1 Optimization of the number of steps at fixed accelerator power

In the previous section, we considered a situation in which the accelerator power could be varied without constraints, allowing its optimum value to be determined in terms of maximizing the brightness and intensity of the neutron beam emerging from the staircase cold source. In particular, the optimum power was found to be approximately 2–2.5 MW. In this section, we consider the case in which the total accelerator power is limited to lower values.

Fig. 14 shows the brightness and intensity gains provided by the staircase assembly as functions of the number of steps $N$, relative to a single moderator with the same overall external dimensions, for different incident proton-beam powers $P$ corresponding to different total moderator lengths $L$ given by $L$=100 $\cdot$ $P$ (with $P$ in MW). These curves are obtained as cross-sections of the 3D surfaces shown in Fig. 11 at fixed accelerator powers.

At $P$ = 0.5 MW, the maximum gains of approximately 1.4, 4.1, and 5.3 are achieved by staircase assemblies comprising 3 steps for $E_{out}$= 3 meV, 9 steps for $E_{out}$= 3 meV, and 12 steps for $E_{out}$= 82 meV, respectively.

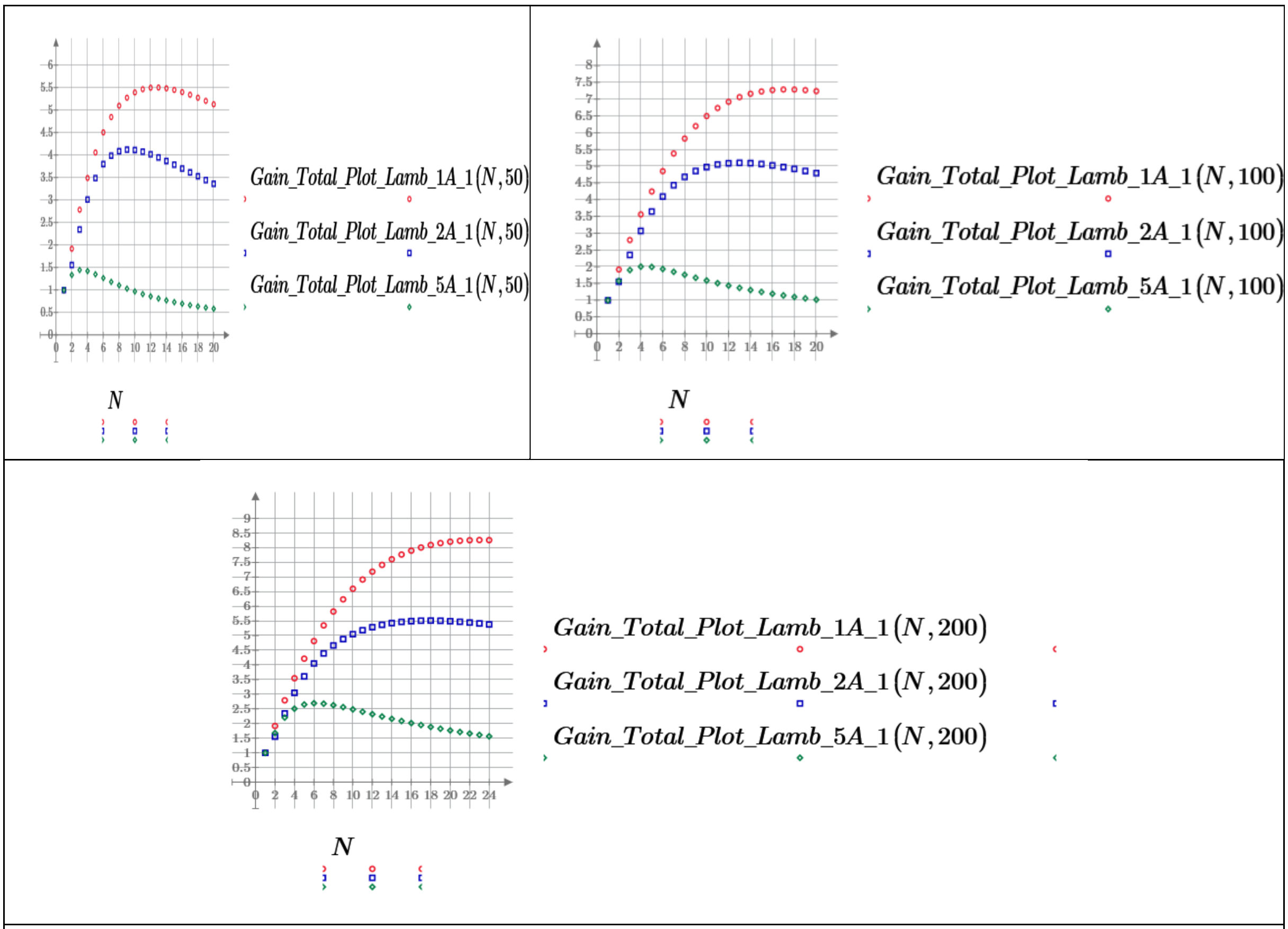

Fig. 14. Brightness gains $G_B^{Stair}$, which are nearly equal to the corresponding intensity gains $G_I^{Stair}$ , provided by the staircase assembly as functions of the number of steps $N$ for neutrons with energies of 3 meV (green circles), 20 meV (blue squares), and 82 meV (red diamonds), at proton-beam powers of (a) $P$ = 0.5 MW ($L$ = 50 cm ), (b) $P$ = 1 MW ($L$ = 100 cm), and (c) $P$ = 2 MW ($L$ = 200 см).

The gain increases with source power, i.e. with the overall length of the moderator. This can be explained by the fact that the optimal number of steps increases less than proportionally with the moderator length. As a result, the optimal length of each individual step also increases: approximately 17 cm for a three-step staircase at 0.5 MW (Fig.14a), 25 cm for a four-step staircase at 1 MW (Fig.14b) and approximately 33 cm for a six-step staircase at 2 MW (Fig.14c). The increase in step length enhances the cumulative build-up of intensity $J(E_{out}, L)$ of the cold-neutron beam emitted from the moderator along its longitudinal axis (see Eq. (3)). However, for 3 meV neutrons, a step length of 17 cm is already large enough to provide about 41% of the maximum achievable intensity, which explains the relatively modest, though still significant, gain of 1.4 obtained with the staircase assembly at a source power of 0.5 MW.

At higher neutron energies, the shorter neutron mean free path leads to a shorter optimal step length and, consequently, a larger optimal number of steps. This further increases the total neutron beam intensity, with the intensity gain reaching values of up to 5 (Figs.10b and 10c).

At $P$ = 1 MW, the maximum gains are significantly higher, reaching approximately 2, 5.1, and 7.2 for staircase assemblies consisting of 4 steps for $E_{out}$ = 3 meV, 12 steps for $E_{out}$ = 20 meV, and 15 steps for $E_{out}$ = 82 meV, respectively. At $P$ = 2 MW, the maximum gains increase further, reaching approximately 2.7, 5.5, and 8.3 for staircase assemblies consisting of 6 steps for $E_{out}$ = 3 meV, 16 steps for $E_{out}$ = 20 meV, and 22 steps for $E_{out}$ = 82 meV, respectively. However, this increase is not proportional to the increase in accelerator power (as also shown in Section 4.1). Using an accelerator with a power above 2.5 MW for a staircase assembly with a total height of $H$=10 cm is inefficient.

### 4.2.2 Trade-off in the choice of the number of steps

However, the maxima of the gain curves as functions of the number of steps $N$ occur at different values of $N$ for different neutron energies. Therefore, selecting the optimum number of steps involves a trade-off. This choice is facilitated by the fact that, for higher-energy neutrons, the maximum occurs over a relatively broad range of $N$.

For example, at a power of 0.5 MW, a three-step staircase assembly provides a maximum gain of 1.4 for 3-meV neutrons, but a relatively modest gain of 2.3 for 20-meV neutrons compared with the maximum achievable value (Fig. 15a). Increasing the number of steps to four reduces the gain for 3-meV neutrons by only 2%, while increasing the gain for 20-meV neutrons to 3. Thus, a 2% reduction in the gain for 3-meV neutrons is accompanied by an increase of approximately 30% for 20-meV neutrons. A similar trade-off is observed at higher accelerator powers of 1 and 2 MW, as shown in Figs. 15b and 15c, respectively.

Thus, the final choice of the number of steps is determined not only by the maximum achievable gain but also by practical considerations related to implementation.

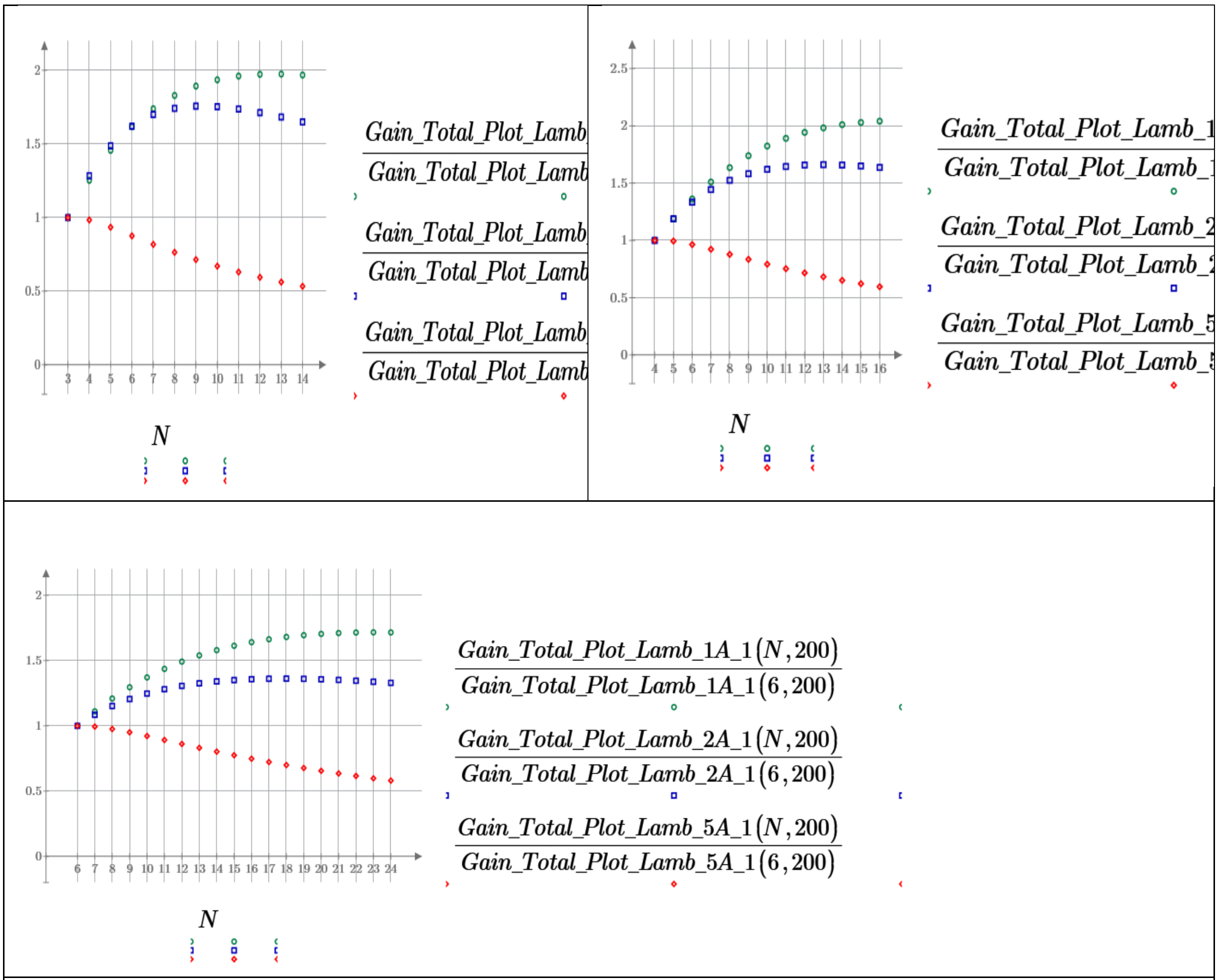

**Fig. 15.** Brightness and intensity gains provided by the staircase assembly, relative to their respective maximum achievable values, as functions of the deviation of the number of steps $N$ from its optimum value. For 3-meV neutrons, the optimum values of $N$ are 3, 4, and 6 at proton-beam powers of (a) $P$ = 0.5 MW ($L$ = 50 cm), (б) $P$ = 1 MW ($L$ = 100 cm) и (c) $P$ = 2 MW ($L$ = 200 cm). Neutron energies are 3 meV (green circles) and 82 meV (red diamonds).

It is important to note that, unlike a single compressed para-hydrogen moderator, for which increasing the accelerator power mainly increases the brightness of the emerging neutron beam in the low-energy range (below 10 meV), the staircase assembly provides gains in both brightness and intensity also for neutrons with energies above 10 meV, emerging in the same direction.

Thus, the staircase assembly represents a natural bi-spectral neutron source: the thermal component is present in the emerging cold-neutron beam, with both components emitted in the same direction, without the use of any neutron-optical devices such as mirror-based bi-spectral switches (e.g., [15], [16]). Unlike the staircase assembly, such devices combine cold neutrons with a thermal neutron beam passing through them and require separate thermal and cold moderators located next to each other.

However, a comparison of the efficiency of these two types of bi-spectral neutron sources is beyond the scope of the present work and will be carried out separately in the future.

## 5. Realistic model of the neutron source

In the previous Monte Carlo calculations, a simplified neutron source model was used with the following approximations: a two-dimensional source of thermal neutrons emulating the premoderator surface was placed directly on the moderator surface, and the tally registered neutrons that had scattered in the moderator exactly once (the single-collision approximation). In reality, the premoderator is located at some distance from the moderator, and neutrons may undergo an unlimited number of collisions in the moderator. Accordingly, in the subsequent simulations, we introduced a gap of 0.7 cm between the premoderator and the moderator, representing a realistic configuration, and removed the restriction on the number of neutron collisions in the moderator.

In addition, real neutron sources employ a reflector to increase the neutron flux in the moderator region. In our simulations, we use a 50-cm-thick beryllium reflector surrounding the moderator on all sides except the neutron exit surface. A comparison of the simulation results obtained with the simplified and realistic models is presented in Fig. 16.

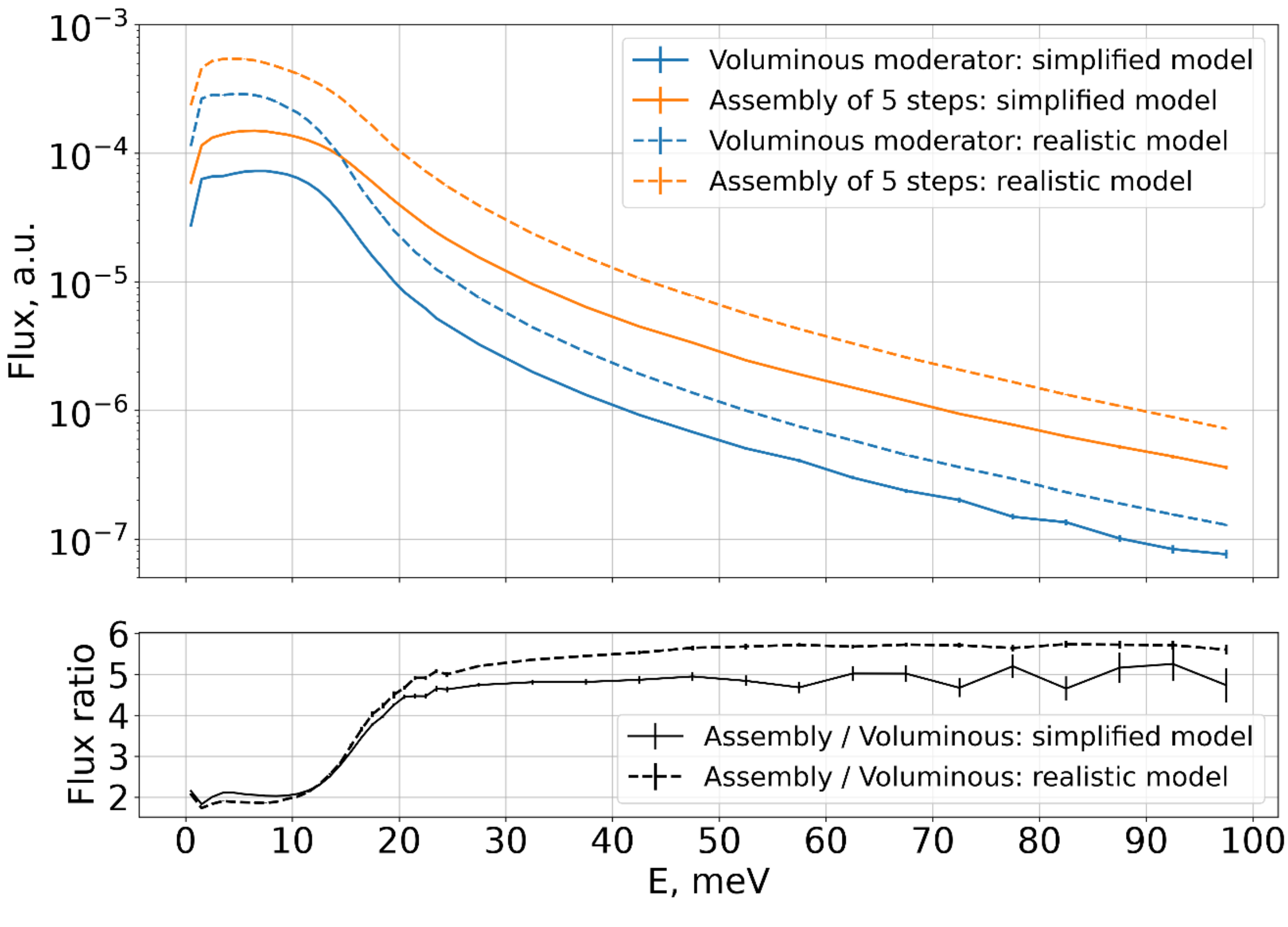

Fig. 16. Top: Neutron spectra of the five-step staircase cold-neutron source (orange) and the voluminous moderator (blue), calculated using PHITS for the simplified model (solid lines) and the realistic model (dashed lines). Bottom: Ratios of the staircase-source spectra to the corresponding voluminous-moderator spectra for the simplified (solid lines) and realistic (dashed lines) models.

### 5.1. Single inclined target-premoderator unit

A simplified version of the proposed geometry of TP units consists in combining the adapted TP units into a single inclined TP unit, as shown in Fig. 17. In principle, this makes it possible to simplify the design of the target; however, such a combination leads to an increase in the distance between thermal pre-moderator and para-hydrogen moderator steps. As a result, the illumination of the latter becomes non-uniform, which leads to a decrease in the intensity of the emitted cold neutron beam.

The losses can be estimated by calculating the degree of the moderator illumination using the geometric view factor $F$, assuming Lambertian (cosine-law) neutron emission from the premoderator [17]:

$$F = \frac{1}{\pi S_{\mathrm{T}}} \int_{S_{\mathrm{T}}} \int_{S_{\mathrm{M}}} \frac{\cos(\theta_{\mathrm{T}}) \cos(\theta_{\mathrm{M}})}{r^2} dS_{\mathrm{M}} dS_{\mathrm{T}} \tag{14}$$

Here $S_M$ and $S_T$ are the surface areas of the moderator and the target, respectively, $r$ is the distance between the corresponding surface elements, and $\theta_M$ and $\theta_T$ are the angles between the neutron propagation direction and the normals to the corresponding surfaces.

By comparing the view factors for the indicated configurations with the view factors for the corresponding reference configurations shown in Fig. 17b, one can calculate the average view-factor reduction coefficient $G_s(N)$ for different numbers $N$ of steps in the assembly (see Appendix 2 for details).

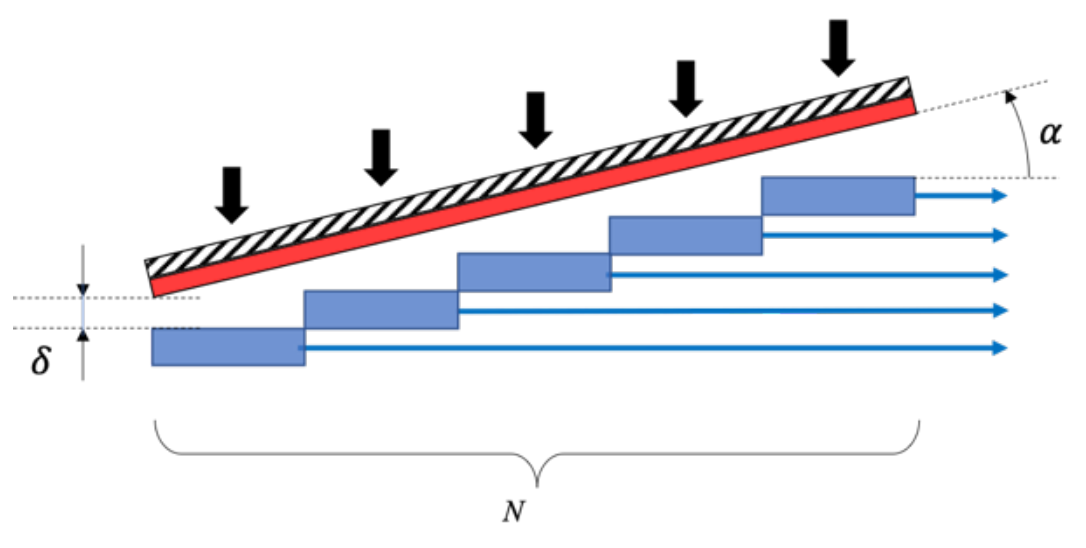

Fig. 17a. A single inclined TP unit for the staircase assembly.

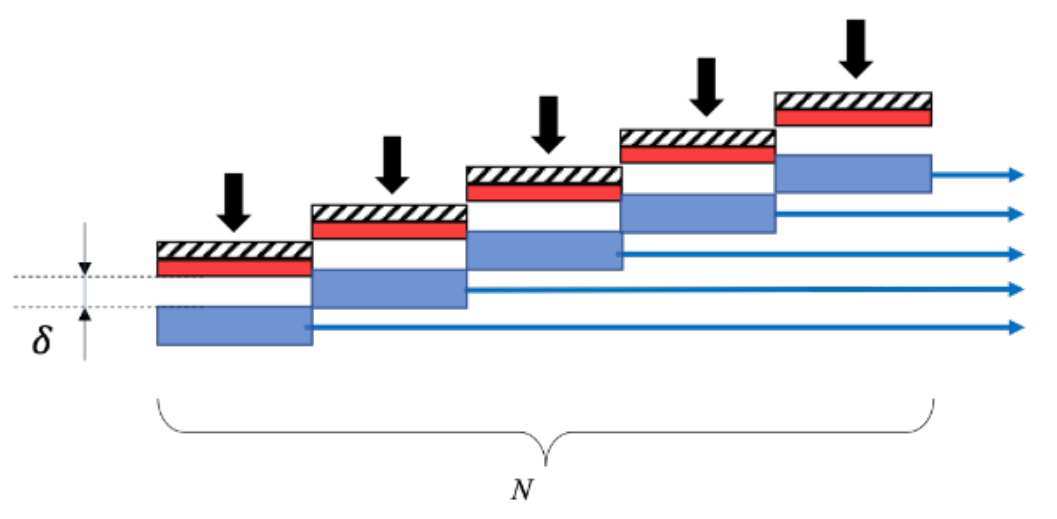

Fig. **17**b. Adapted TP units for the staircase assembly.

The calculations were performed by numerical integration, and the results are presented in Fig. 18. Combining the adapted TP units into a single inclined TP unit leads to a decrease in the average geometric illumination of the p-$H_2$ moderators of about 7% for a linear assembly of 5 steps (as shown in Fig. 17a), which results in the corresponding decrease in the intensity of the emitted cold neutron beams is of the same magnitude. The tendency of the average view-factor reduction coefficient $G_s(N)$ to saturate with increasing $N$ in Fig. 18 is explained by the decreasing contribution of the peripheral steps, which see a smaller source area than the central steps.

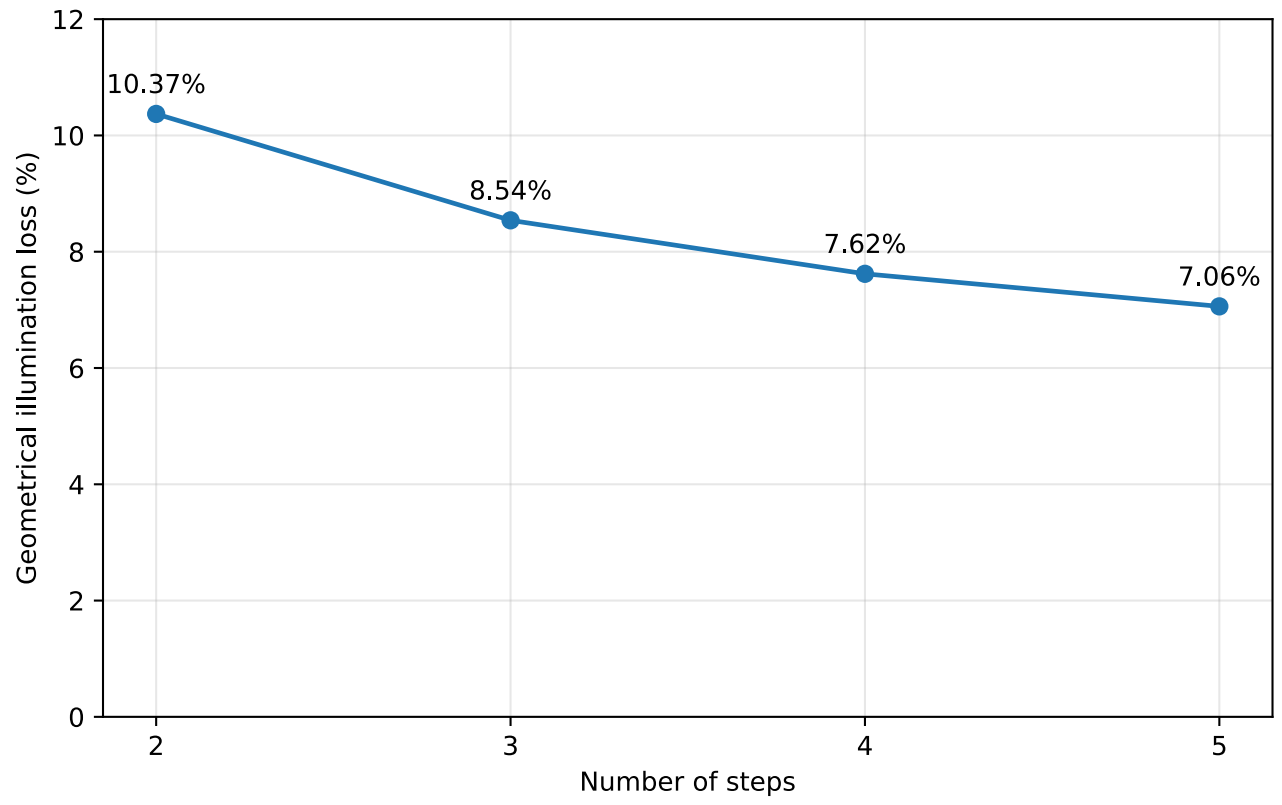

Fig. **18**. Dependence of the average view-factor reduction coefficient $F_{incl}$ on the number of steps $N$ for assemblies with a total length of $L$ =50 cm. The minimum distance between a target unit and the moderator surface is $\delta$ = 0.7 cm.

Fig. 19 presents the results of PHITS simulations for the 5-step staircase cold neutron source with a single inclined TP unit and with adapted TP units (Fig. 17a and Fig. 17b respectively). These results in principle confirm the estimate presented in Fig. 18.

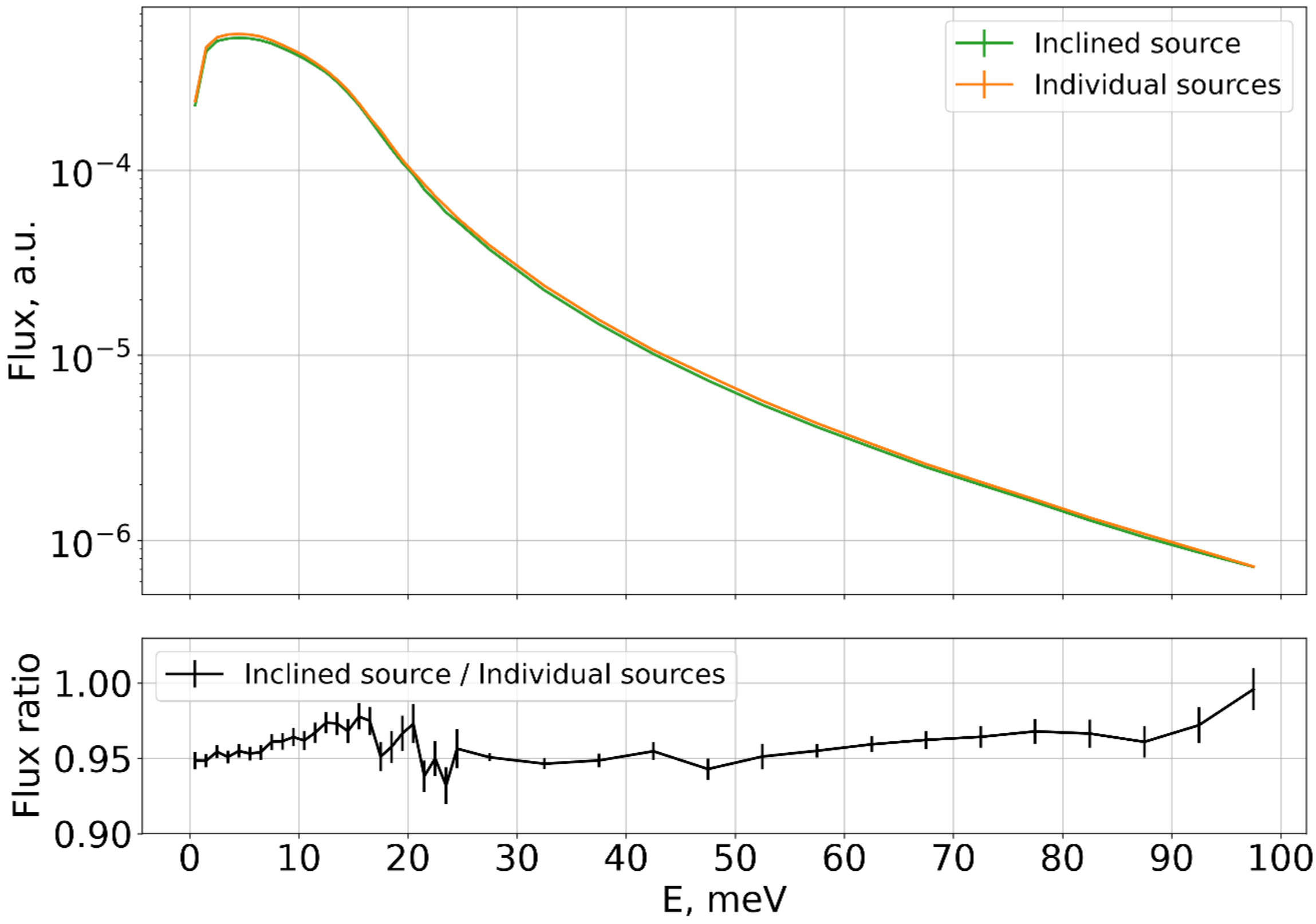

Fig. 19. Top: spectra of the 5-step staircase cold neutron source with five adapted TP units (orange) and with a single inclined TP unit (blue), calculated with PHITS. Bottom: ratio of these spectra.

## 5.2 V-Shaped configuration

Despite the high efficiency of the staircase cold source, its practical implementation may be complicated by the considerable longitudinal length of the assembly. As we noted earlier [8], this drawback can be substantially reduced by using a V-shaped configuration of the staircase cold source, in which one half of the steps are illuminated from one side and the other half from the opposite side (Fig. 20). In this case, the longitudinal dimension of the assembly is reduced by a factor of two for an even number of steps and by a somewhat smaller amount for an odd number of steps.

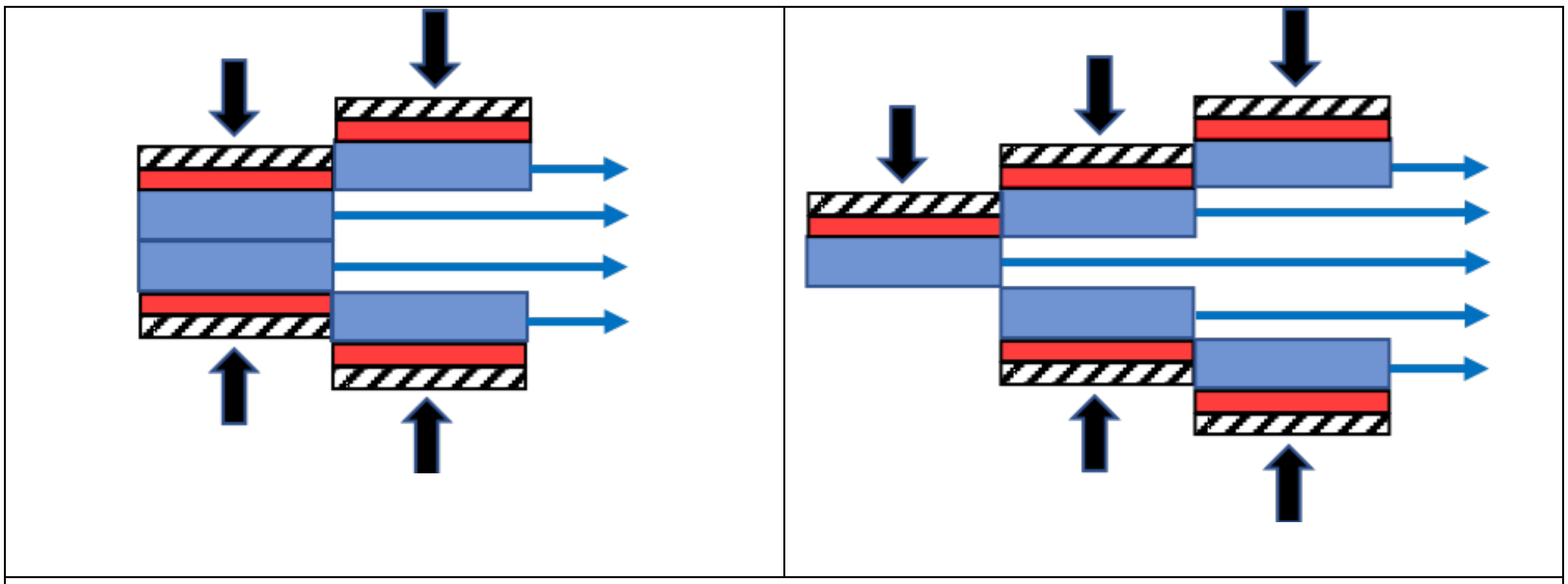

Fig. **20**. Variants of the V-shaped configuration for four-step and five-step staircase assemblies.

Since the geometrical parameters of the individual steps and their local illumination conditions remain the same as in the linear staircase assembly considered above, the brightness and intensity gains remain unchanged. At the same time, the V-shaped configuration may offer additional advantages.

First, in a linear staircase assembly, due to shielding by the target and premoderator (Fig. 5b), cold neutrons are emitted to the left from only one step. Consequently, the brightness of this beam corresponds to that of a compressed moderator with a thickness of 2 or 2.5 cm, but its intensity is lower by a factor of 4–5. In the V-shaped configuration, cold neutrons are emitted to the left from two steps. Thus, the brightness of the beam again corresponds to that of a compressed moderator with a thickness of 2 or 2.5 cm, but its intensity is lower by only a factor of 2–2.5.

Second, in a linear staircase assembly, each step is illuminated only by thermal neutrons from one side. In the V-shaped configuration, however, a step can also receive thermal neutrons from the opposite side. Therefore, an additional increase in the illumination of the steps and, consequently, a small additional gain in intensity can be expected. This assumption is illustrated in Fig. 21: the Monte Carlo simulation results show an intensity that is 10–15% higher than that predicted by the analytical calculations.

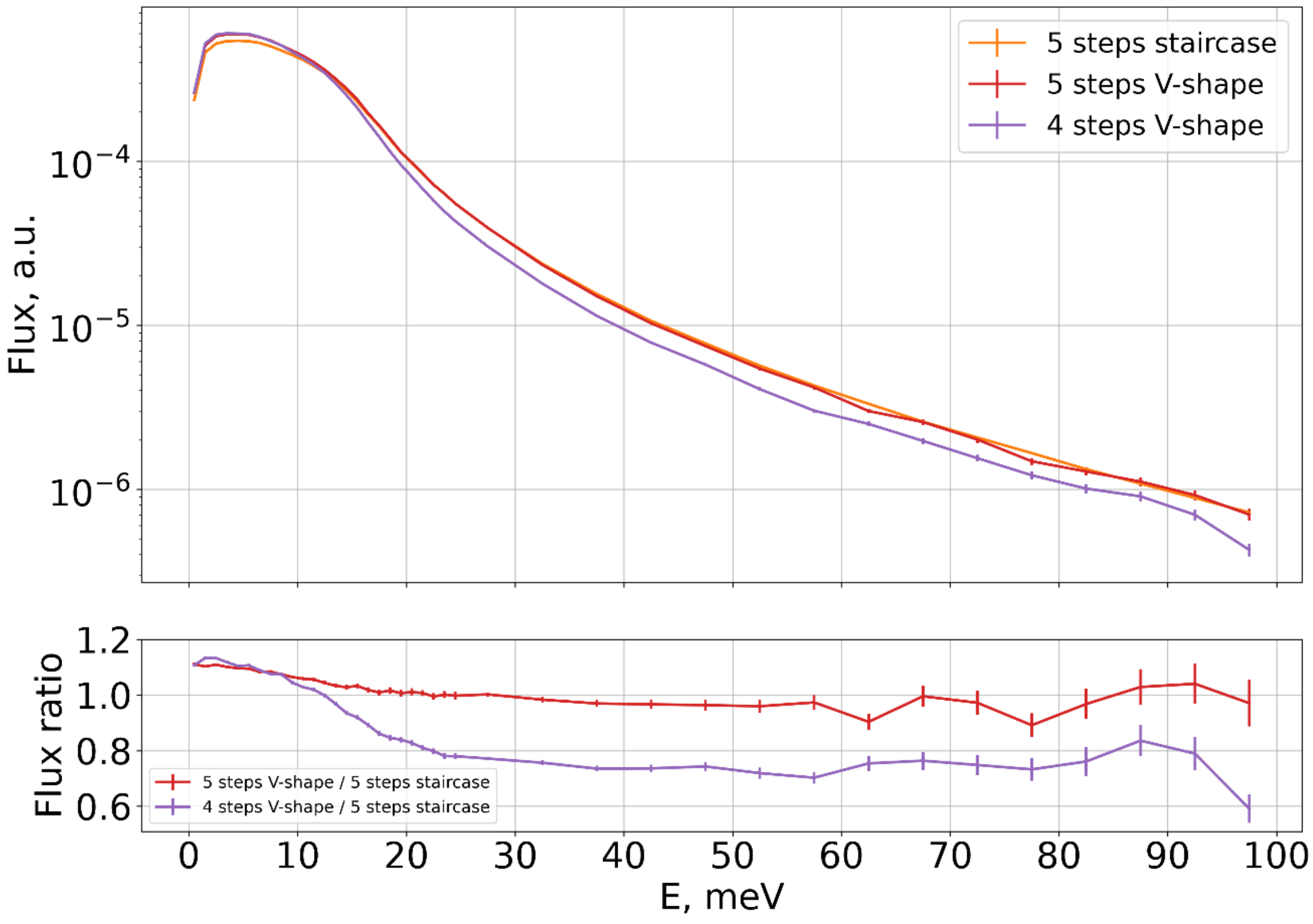

Fig. 21. Top: Neutron spectra of the 5-step staircase cold neutron source (orange), the 5-step V-shaped moderator (red), and the 4-step V-shaped moderator (violet), calculated using PHITS. Bottom: Ratios of the spectra of the 5-step (red) and 4-step (violet) V-shaped moderators to that of the staircase cold neutron source.

### 5.2.1 Beam angular distributions

The moderator geometries considered here exhibit substantially different angular distributions of neutrons emerging from their exit surfaces. These distributions, calculated using PHITS, are presented in Fig. 22. The realistic model described above was used in the calculations.

A point detector was sequentially rotated in the horizontal plane ($L \times W$) and the vertical plane ($L \times H$) around the center of the moderator exit surface ($10 \times 10$ cm²), registering neutrons with energies up to 100 meV. To ensure that only neutrons emerging from the moderator exit surface were registered, the portion of the beryllium reflector extending beyond the plane of this surface toward the detector was removed, while the remaining part was covered with an opaque shield containing a window of the corresponding area $10 \times 10$ cm².

As can be seen from Fig. 22, the vertical angular distributions of the voluminous and staircase cold sources are asymmetric, in contrast to those of the V-shaped moderators. This is because the former are illuminated by thermal neutrons from only one side, whereas in the latter are illuminated from two sides. In addition, during vertical scanning of the staircase cold sources, individual steps successively move out of the detector's field of view, giving rise to characteristic kinks in the angular distribution curves.

It is worth noting that, for the V-shaped moderators, the horizontal and vertical angular distributions are very similar to each other (in contrast, for example, to those of single compressed moderators). This makes V-shaped moderators equally efficient for use in instruments sensitive to both the vertical and horizontal divergence of the neutron beam.

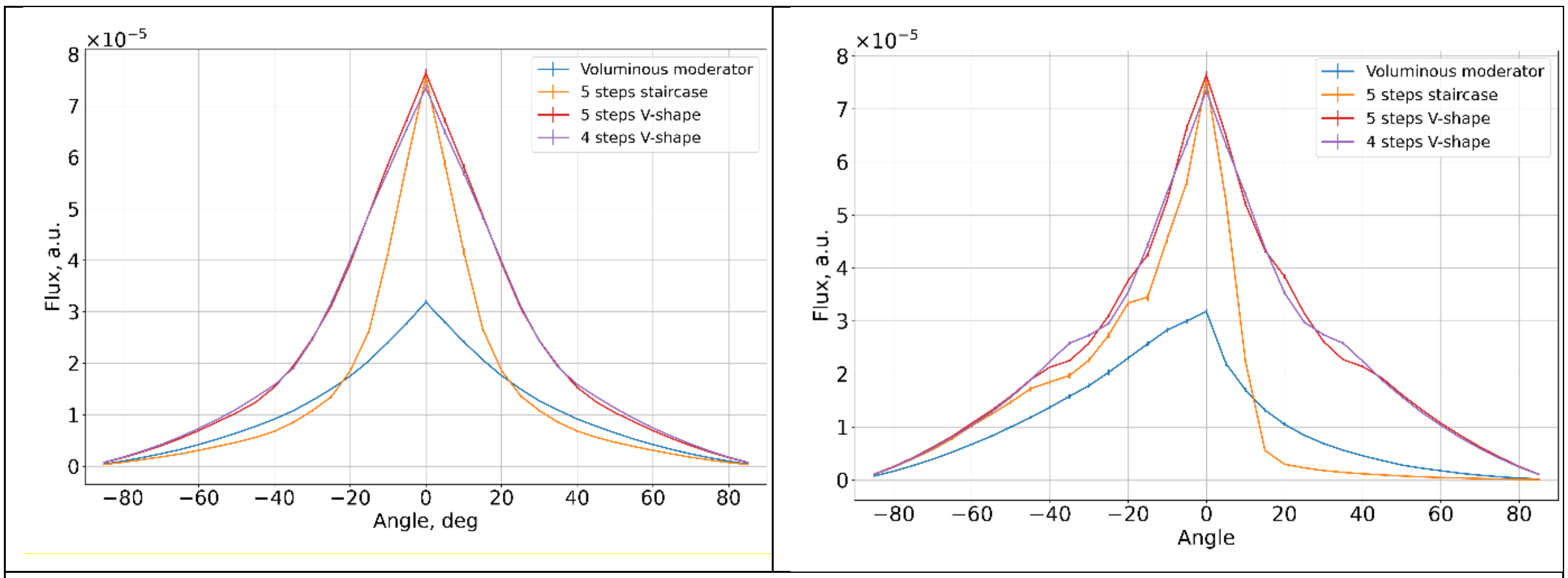

Fig. 22. Horizontal (left) and vertical (right) angular distributions of neutrons emerging from the moderator exit surface calculated using PHITS. Blue curves – voluminous moderator; orange curves – 5-step staircase cold neutron source; red curves – 5-step V-shaped moderator; violet curves – 4-step V-shaped moderator.

## 6. Conclusion

The main result of this work is the demonstration that a further increase in the efficiency of low-dimensional para-hydrogen moderators requires optimization not only of the geometry of the cryogenic moderator itself but also of the geometry of the thermal neutron source.

It is shown that a simple increase in the moderator length, accompanied by a proportional increase in the dimensions of the thermal neutron source and in the accelerator power, is an ineffective way to further increase the intensity. For a 50-cm-long moderator, the efficiency of utilizing additional accelerator power is about 35% for neutrons with a wavelength of 5 Å and only about 7% for neutrons with a wavelength of 2 Å. Thus, a further increase in accelerator power cannot be considered independently of the geometry of the system that converts this power into a flux of thermal and then cold neutrons.

To overcome this limitation, the work proposes using a staircase assembly [8] consisting of several short compressed pH2 moderators, each illuminated by its own target–premoderator unit.

Illuminating each individual moderator with its own target–premoderator unit makes it possible to avoid under-illumination of the staircase assembly that occurs when using a conventional “target–thermal premoderator” configuration. In such a system, the limited dimensions of the thermal-neutron generation region, together with the different distances of individual steps from the thermal premoderator, lead to non-uniform illumination of the steps and, consequently, to a substantial decrease in the efficiency of the staircase assembly.

Splitting a long moderator into several short, compressed elements and adapting the target–premoderator system accordingly make it possible to use the available accelerator power more efficiently while simultaneously increasing the brightness and intensity of the emerging neutron beam. For the same external dimensions of the staircase assembly and the reference single moderator, the brightness gain obtained by compressing the individual steps is almost entirely translated into a corresponding gain in total intensity. In other words, the staircase assembly makes it possible to avoid the trade-off between increasing brightness and decreasing intensity that is characteristic of a single compressed moderator, thereby preserving the advantages of geometric compression of individual moderators while simultaneously increasing the overall cross-sectional dimensions of the emerging cold-neutron beam.

This facilitates coupling the source to neutron guides located at a distance of approximately 1.5–2 m without substantial under-illumination of their entrance cross-sections, which would otherwise lead to significant losses in the brightness of the neutron beams delivered to the samples. At the same time, the volume of para-hydrogen used, and the corresponding heat deposition in it, are substantially smaller than those for a reference volumetric moderator with the same external dimensions.

The final choice of the number of steps is determined not only by the maximum achievable gain but also by practical constraints. From this point of view, the five-step staircase is particularly attractive, as the dimensions of the individual steps are $(10 \times 10)$ cm$^2$. Compared with the four-step staircase, the gain for 20-meV neutrons increases by approximately 15%, reaching a value of about 4, whereas for cold neutrons with energies of 3–9 meV, the decrease in gain is only 2–4%. Thus, the five-step system represents a successful compromise between efficiency over a broad energy range and practical simplicity of implementation.

A particularly interesting feature of the staircase cold source is the change in the spectral characteristics of the emerging beam. For the five-step system, the intensity gain begins to increase rapidly at energies above approximately 10 meV and reaches about 4.5 at energies above 25 meV, remaining at this level over a broad range of thermal-neutron energies. Thus, unlike a single compressed moderator, whose main advantage manifests itself predominantly in the low-energy region, the staircase assembly also provides a significant gain for neutrons with energies above 10–20 meV. In this sense, the proposed system possesses the characteristics of a neutron source with an extended energy range.

For a system with a total length of 50 cm, consisting of five steps and illuminated by a proton beam with five times the power, a substantial gain is obtained compared with a single moderator of the same external dimensions, ($(10 \times 10 \times 50)$ cm$^3$, illuminated by a proton beam of the same total power. For cold neutrons with ($\lambda > 2.6$) Å, the intensity increases by approximately a factor of 1.5, whereas for neutrons with energies above 20 meV, the gain reaches approximately 4.6. Thus, the proposed geometry makes it possible to utilize the increased accelerator power significantly more efficiently.

The proposed V-shaped configuration provides an additional possibility for practical optimization. By illuminating the two halves of the staircase from opposite sides, it reduces the longitudinal dimensions of the assembly by a factor of 2–2.5 without compromising its main neutronic characteristics. Combining the individual target assemblies into a single continuous inclined target–premoderator system, which may simplify the technical implementation, reduces the geometrical illumination of the individual steps and results in a gain reduction of approximately 7–10% for the linear configuration. This reduction is considerably smaller than the potential gain provided by the staircase assembly.

The results obtained can be used in the development of both compact accelerator-driven neutron sources and more powerful spallation sources with stationary targets, for which further increases in accelerator power are limited by the maximum permissible heat load on the target. In such systems, the use of several targets geometrically adapted to the staircase cold source can provide a more efficient distribution of the thermal load while simultaneously increasing the intensity of both thermal and cold neutrons emitted by the p-$H_2$ cold-neutron source, thereby improving accelerator-power utilization.

**Appendix 1.**

In this work, we restrict ourselves to a two-dimensional treatment: from various points of the premoderator located at a distance $\Delta$ from its edge, thermal neutrons are emitted at an angle $\theta$ in the $xz$ half-plane perpendicular to the moderator width, see Fig. A1. For angles $0 < \theta < \pi/2$, the path length $s$ until exit from the moderator of height $H$ and length $L$ can be written as:

$$S(\theta, H, L, \Delta) = \begin{cases} \frac{H}{\cos(\theta)}, & \theta < \theta_0 \\ \frac{L-\Delta}{\sin(\theta)}, & \theta \geq \theta_0 \end{cases} \qquad \text{(A1.1)}$$

where $\theta_0 = \arctan((L - \Delta)/H)$ is the angle between the diagonal and the height $H$, see Fig. A1. To simplify the calculations, we use the average path length of trajectories emitted from different points of the premoderator at angles $-\pi/2 < \theta < \pi/2$:

$$S(\theta, H, L) = 2 \cdot \frac{1}{L} \int_0^L s(\theta, H, L, \Delta)\, d\Delta \qquad \text{(A1.2)}$$

Here, emission at angles $-\pi/2 < \theta < 0$ can be accounted for by a factor of 2 owing to the symmetry of the problem.

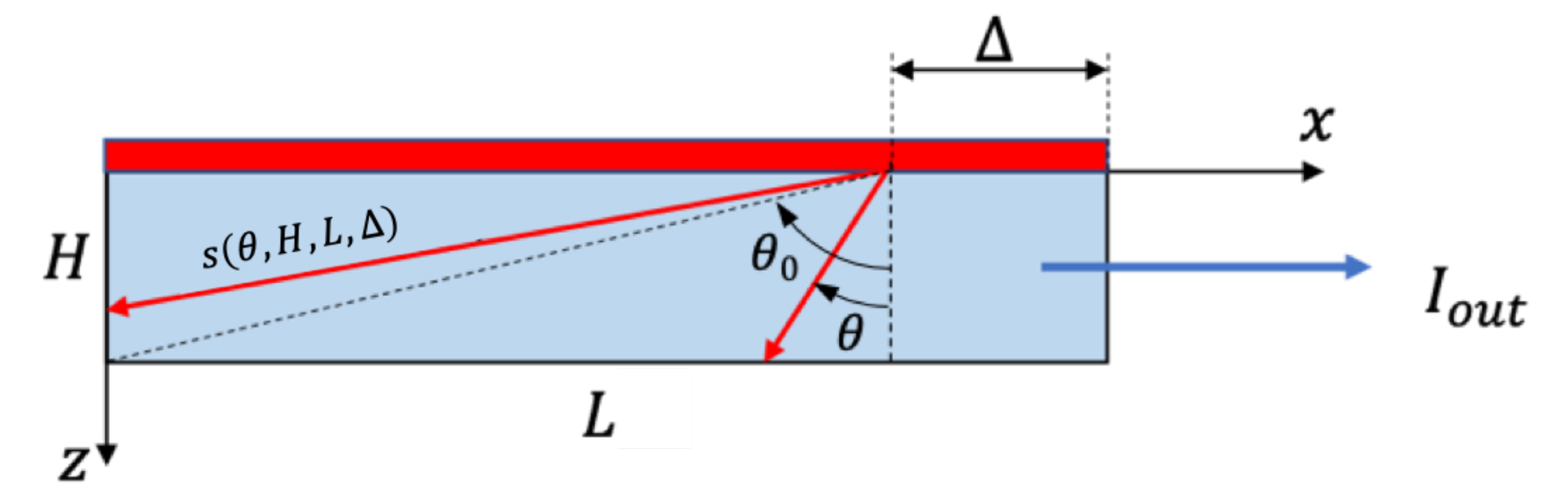

Fig. A1. Trajectories of thermal neutrons emitted from different points of the premoderator.

**Appendix 2**

Consider a configuration consisting of a single continuous inclined target-premoderator unit of length $L$ and a staircase cold source assembly of $N$ parahydrogen moderators of length $L_0 = L/N$ and height $h$ (all elements having width $W$), as shown in Fig. 17 and reproduced here in Fig. A2 for the reader's convenience. The inclination angle of the target relative to the moderator surface is $\alpha = \arctan(h/L_0)$, and the minimum distance between the target unit and the moderator surface is $\delta$.

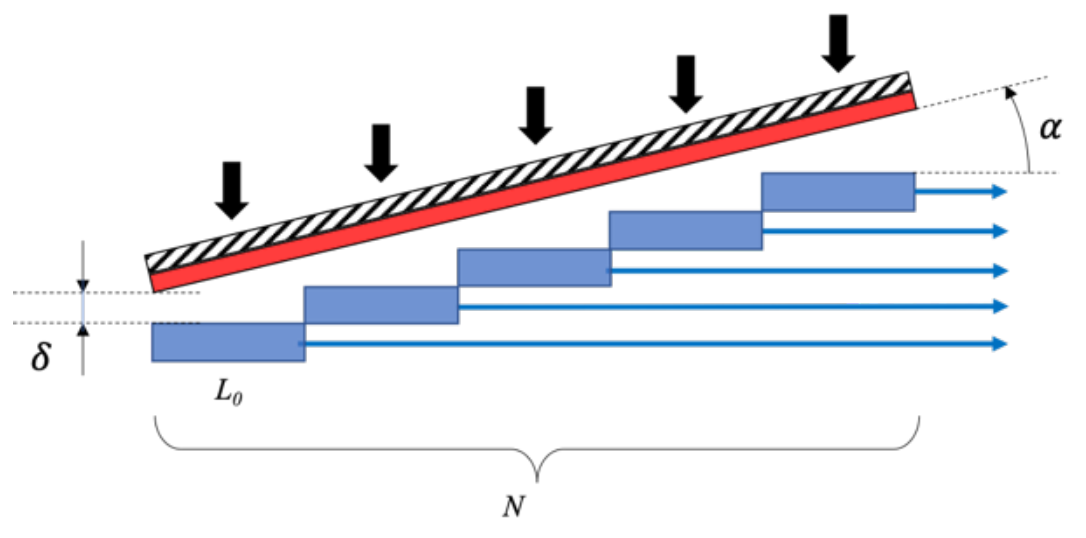

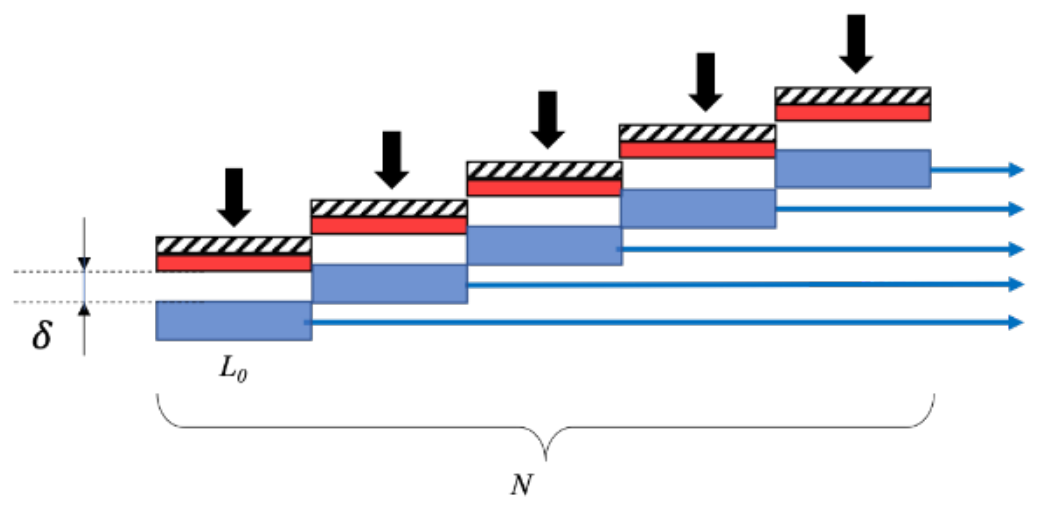

| Fig. A2a. A single inclined target–premoderator unit for the staircase assembly. | Fig. A2b. Adapted target-premoderator units for the staircase assembly. |
|---|---|

The illumination $F_i$ of the surface of each step of the staircase cold source can be estimated using the geometric view factor (see Eq. (14)). When calculating the illumination of each moderator, the contribution of the entire surface of the target is taken into account, not only the TP unit located directly above the moderator. The average view factors are calculated as

$$F_{mean} = \frac{1}{N}\sum_{i=1}^{N} F_i$$

As a reference configuration, $N$ independent TP units of length $L_0$ are considered, each positioned parallel to the corresponding para-hydrogen moderator at a distance $\delta$. $F_{ref}$ is the view factor for one step of the reference moderator. Neglecting additional illumination of the end faces of the para-hydrogen moderator steps by thermal neutrons arriving from adjacent water premoderators, all steps in the reference moderator are equivalent, $F_{ref}$ is the view factor for the entire reference moderator. The average loss factor of single inclined TP unit relative to the reference configuration is defined as

$$G_S = 1 - \frac{F_{mean}}{F_{ref}}$$

For a step of length $L_0 = 10$ cm and height $h = 2$ cm, which corresponds to the case $N = 5$ and $L = 50$ cm considered above in this article, the angle is $\alpha = 11.3°$. The minimum distance between the target unit and the moderator surface is assumed to be $\delta = 0.7$ cm.

The view factor values for individual moderators calculated by numerical integration, as well as the average view factors and loss factors, are presented in Table 1. $F_{ref}$ for one step of the reference moderator is 0.874. The numerical convergence of the calculation was verified by successively increasing the number of integration grid nodes. The change in the calculated loss value when increasing the number of nodes from 300 to 600 does not exceed 0.00005%, so the numerical integration error is substantially smaller than the accuracy of the geometric parameters and can be considered negligible.

Tab. 1.

| Number of steps, $N$ | $F_1, \ldots, F_5$ | $F_{mean}$ | $G_S$ |
|---|---|---|---|
| 2 | 0.801, 0.767 | 0.78 | 10.3 % |
| 3 | 0.80, 0.83, 0.77 | 0.80 | 8.0 % |
| 4 | 0.80, 0.83, 0.83, 0.77 | 0.81 | 7.0 % |
| 5 | 0.80, 0.83, 0.83, 0.83, 0.77 | 0.81 | 7.0 % |

Thus, combining two separate target assemblies into a single inclined target leads to a reduction in the average geometric illumination of the moderators by approximately (7–10) %.

As the number of the steps increases from two to three, $F_{mean}$ increases, while the relative loss of geometric illumination $G_S$ decreases owing to the reduced relative contribution of the peripheral, less well-illuminated steps. With a further increase in the number of steps, this illumination loss becomes insignificant, and $F_{mean}$ and $G_S$ remain essentially unchanged. Consequently, in the geometry under consideration, the magnitude of the loss is determined mainly by the local inclination of the target and the distance between the target and the moderators, rather than by the total length of the target assembly.

It should be noted that the actual length of the inclined target in Fig. 17a or A2a is $L/\cos(\alpha)$; however, for inclination angles $\alpha$ of 11° or 14°, the correction amounts to only 2–3% and is not considered here.

**Acknowledgement**

This work has been supported by the Talent Introduction Program of the Institute of High Energy Physics No. E6515CU110 "Physical design of the staircase-style high brightness cold neutron moderator for the accelerator-based neutron source"

**References**

**1.** Andersen, Ken Holst, et al. "Optimization of moderators and beam extraction at the ESS." *Applied Crystallography* 51.2 (2018): 264-281.

2. Batkov, Konstantin, et al. "Unperturbed moderator brightness in pulsed neutron sources." *Nuclear Instruments and Methods in Physics Research Section A: Accelerators, Spectrometers, Detectors and Associated Equipment* 729 (2013): 500-505.

3. Zanini, L., et al. "Design of the cold and thermal neutron moderators for the European Spallation Source." *Nuclear Instruments and Methods in Physics Research Section A: Accelerators, Spectrometers, Detectors and Associated Equipment* 925 (2019): 33-52.

4. Gallmeier, Franz X., and Igor Remec. "A liquid hydrogen tube moderator arrangement for SNS second target station." *Review of Scientific Instruments* 93.8 (2022).

5. Cronert, Tobias, et al. "High brilliant thermal and cold moderator for the HBS neutron source project Jülich." *Journal of Physics: Conference Series*. Vol. 746. No. 1. IOP Publishing, 2016.

6. Konik, Petr, and Alexander Ioffe. "A new method to find out the optimal neutron moderator size based on neutron scattering instrument parameters." *Nuclear Instruments and Methods in Physics Research Section A: Accelerators, Spectrometers, Detectors and Associated Equipment* 1056 (2023): 168643.

7. Baggemann, J., et al. "High power target for the high brilliance neutron source." *Nuclear Instruments and Methods in Physics Research Section A: Accelerators, Spectrometers, Detectors and Associated Equipment* 1069 (2024): 169912.

8. Ioffe, Alexander, Petr Konik, and Konstantin Batkov. "A concept of a para-hydrogen-based cold neutron source for simultaneous high flux and high brightness." *Journal of nuclear engineering* 6.1 (2025): 3.

9. Mezei, Ferenc, et al. "Low dimensional neutron moderators for enhanced source brightness." *Journal of Neutron Research*17.2 (2014): 101-105.

10. Watanabe, Noboru. "Neutronics of pulsed spallation neutron sources." *Reports on Progress in Physics* 66.3 (2003): 339-381.

11. Gutberlet, Thomas, et al. "High-Current Accelerator-Driven Neutron Sources (HiCANSs): Current Projects and Perspectives in Europe." *Quantum beam science* 10.4 (2026): 26.

12. Shapiro, Dmitrii, et al. "New para-hydrogen cold neutron source at the Budapest Research Reactor: Monte Carlo simulations." *Applied Crystallography* 59.2 (2026).

13. Sato, Tatsuhiko, et al. "Recent improvements of the particle and heavy ion transport code system–PHITS version 3.33." *Journal of Nuclear Science and Technology* 61.1 (2024): 127-135.

14. Granada, J. R. "Slow neutron total cross-section of Al6061 at low temperatures." *Journal of nuclear materials* 277.2-3 (2000): 346-350.

15. Krist, Th, and A. Tennant. "Guide and Instrument Upgrade at Helmholtz-Zentrum Berlin." *Neutron News* 25.2 (2014): 4-7.

16. Schweika, Werner, et al. "DREAM—a versatile powder diffractometer at the ESS." *Journal of Physics: Conference Series*. Vol. 746. No. 1. IOP Publishing, 2016.

17. F. P. Incropera, D. P. DeWitt, T. L. Bergman, A. S. Lavine, *Fundamentals of Heat and Mass Transfer*, 6th ed., John Wiley & Sons, 2007, Section 13.1.1, Eq. (13.1).